\documentclass[11pt,a4paper]{article}
\pdfoutput=1 
             
\usepackage{graphicx}
             
\usepackage[utf8x]{inputenc}	
\usepackage[english]{babel}
\usepackage{paralist}
\usepackage{amsmath}
\numberwithin{equation}{section}
\usepackage{amssymb}
\usepackage{amsthm}
\usepackage{MnSymbol}
\usepackage{mathtools}		
\usepackage{dsfont}		
\usepackage{stmaryrd}		
\usepackage{cite}		
\usepackage[svgnames]{xcolor}
\usepackage{enumerate}
\usepackage{setspace}
\usepackage[normalem]{ulem}

\usepackage{booktabs}
\usepackage{tikz}
\usetikzlibrary{shapes}
\usetikzlibrary{positioning}
\usetikzlibrary{chains}
\usetikzlibrary{arrows,fit,decorations.pathreplacing, shapes.misc}
\usetikzlibrary{decorations.markings}
\tikzstyle{every picture}+=[remember picture]
\tikzstyle{na} = [baseline=-.5ex]
\usepackage{todonotes}
\usepackage{tikz-3dplot}
\usepackage{pgfplots}
\usepackage[font=small,labelfont=bf]{caption}
\usepackage[font=small,labelfont=bf]{subcaption}
\usepackage[pdftex,breaklinks,colorlinks=true,urlcolor=RoyalBlue,linkcolor=blue,
citecolor=blue]{hyperref}

\def\mc{\mathcal}
\def\beq{\begin{equation}}
\def\eeq{\end{equation}}
\def\beqn{\begin{eqnarray}}
\def\eeqn{\end{eqnarray}}

\def\half{\frac{1}{2}}

\def\Tr{{\rm Tr}}
\def\X{\!&\times &}
\def\={\!&=&}
\def\+{\!&+&}
\def\-{\!&-&}

\def\for{\ \ \ {\rm for} \ \ }

\newcommand{\none}{${\mathcal N}=1\,\,$}

\newcommand{\noz}{${\mathcal N}=(1,0)\,\,$}

\catcode`@=11
\allowdisplaybreaks

\begin{document}
\begin{titlepage}
\setcounter{page}{0}

\begin{center}

{\Large\bf 
4d $\mathcal{N}=1$ from 6d D-type $\mathcal{N}=(1,0)$
}

\vspace{15mm}

{\large Jin Chen${}^{a}$},\ 
{\large Babak Haghighat${}^{b}$},\ 
{\large Shuwei Liu${}^{c}$},\  and \ 
{\large Marcus Sperling${}^{b}$} 
\\[5mm]
\noindent ${}^{a}${\em CAS Key Laboratory of Theoretical Physics, Institute of 
Theoretical Physics}\\
{\em Chinese Academy of Sciences, Beijing 100190, China}\\
{Email: {\tt jinchen@itp.ac.cn}}
\\[5mm]
\noindent ${}^{b}${\em Yau Mathematical Sciences Center, Tsinghua University}\\
{\em Haidian District, Beijing, 100084, China}\\
{Email: {\tt babakhaghighat@tsinghua.edu.cn},} \\  
{{\tt marcus.sperling@univie.ac.at }}
\\[5mm]
\noindent ${}^{c}${\em Department of Physics, Tsinghua University}\\
{\em Haidian District, Beijing, 100084, China}\\
{Email: {\tt liu-sw15@mails.tsinghua.edu.cn}}
\\[5mm]
\vspace{15mm}

\begin{abstract}
Compactifications of $6$d $\mathcal{N}=(1,0)$ SCFTs give rise to new 
$4$d $\mathcal{N}=1$ SCFTs and shed light on interesting dualities between such 
theories. In this paper we continue exploring this line of research by 
extending 
the class of compactified $6$d theories to the $D$-type case. The simplest such 
$6$d theory arises from D5 branes probing $D$-type singularities. Equivalently, 
this theory can be obtained from an F-theory compactification using $-2$-curves 
intersecting according to a $D$-type quiver. Our approach is two-fold. We start 
by compactifying the $6$d SCFT on a Riemann surface and compute the central 
charges of the resulting $4$d theory by integrating the $6$d anomaly polynomial 
over the Riemann surface. As a second step, in order to find candidate $4$d UV 
Lagrangians, there is an intermediate $5$d theory that serves to construct $4$d 
domain walls. These can be used as building blocks to obtain torus 
compactifications. In contrast to the $A$-type case, the vanishing of anomalies 
in the $4$d theory turns out to be very restrictive and constraints the choices 
of gauge nodes and matter content severely. As a consequence, in this paper one 
has to resort to non-maximal boundary conditions for the $4$d domain walls. 
However, the comparison to the $6$d theory compactified on the Riemann 
surface becomes less tractable.
\end{abstract}

\end{center}

\end{titlepage}

{\baselineskip=12pt
{\footnotesize
\tableofcontents
}
}

%
\section{Introduction}
\label{intro}
Recently, a series of interesting works 
\cite{Gaiotto:2015usa,Razamat:2016dpl,Bah:2017gph,Kim:2017toz,Kim:2018bpg,Kim:2018lfo,Razamat:2018gro}
initiated a systematic study on compactifications of various $6$d \noz SCFTs 
on 
Riemann surfaces to obtain a vast class of new $4$d \none SCFTs. The 
corresponding SCFTs in six and four dimensions are then connected through RG 
flows which preserve certain properties of the $6$d fixed point theory along 
the 
flow. This construction has given rise to new dualities between $4$d 
$\mathcal{N}=1$ theories by tracing different theories back to the same $6$d 
origin \cite{Razamat:2018gro} as well as new asymptotically free UV 
descriptions of $4$d SCFTs. 

In all these examples, one starts with a $6$d theory obtained by 
compactifying F-theory on a local elliptic Calabi-Yau threefold. In this 
construction, as initiated by \cite{Heckman:2013pva}, the geometry of the base 
$B$ of the Calabi-Yau manifold gives rise to the tensor multiplet sector of the 
$6$d SCFT such that the number of tensor multiplets is equal to the dimension 
of 
$H^{1,1}(B,\mathbb{Z})$. Furthermore, the intersection form on $B$ gives the 
couplings of these tensor multiplets to each other. This intersection form is 
constrained by the fact that all curve classes inside the base must be 
simultaneously shrinkable to zero volume in order to restore conformal symmetry 
at the origin of the tensor branch. As a consequence all curve classes are 
forced to be $\mathbb{P}^1$'s which have negative self-intersection number. The 
elliptic fiber above these curves degenerates and gives rise to gauge groups 
determined by Kodaira's classification of elliptic fibers in the effective $6$d 
theory. Following the classification of \cite{Heckman:2013pva}, the 
$\mathbb{P}^1$'s in the base can intersect according to a generalised $A$-type 
or a generalised $D$-type quiver, and in cases where all $\mathbb{P}^1$'s have 
self-intersection number $-2$ one can also construct $E$-type quivers. 

Compactifications of such theories to $4$d $\mathcal{N}=1$ was initiated in 
\cite{Gaiotto:2015usa,Razamat:2016dpl}. Therein, the authors focused on the 
simplest possibility, namely starting with a $6$d theory which arises from an 
$A$-type quiver of $-2$ curves. Already, this simple case gives rise to an 
immensely rich class of $4$d theories admitting asymptotically free UV 
Lagrangian 
descriptions. To obtain the theory on a general Riemann surface, one constructs 
the results for the torus and the three-punctured sphere and all other cases can 
be obtained by gluing these building blocks. A stepping stone for these 
constructions is the theory corresponding to the tube, namely the two-punctured 
sphere. Here, the idea is to first find the circle compactification of the $6$d 
theory 
giving rise to a $5$d gauge theory and subsequently constructing $4$d domain 
walls 
for these $5$d theories by choosing $\half$-BPS boundary conditions. In 
practice, 
this construction leads to $SU(N)$ ($N$ here being the number of 
nodes/$\mathbb{P}^1$'s in the $6$d quiver) gauge nodes forming a tessellation 
of 
the tube corresponding to the two-punctured sphere. It is found that the number 
of the Cartans of the global symmetry group is preserved along the RG flow from 
$6$d to $4$d and manifests itself as $U(1)$ flavour symmetries of the $4$d 
Lagrangian 
theory. Moreover, the $4$d theory has further flavour symmetries which 
correspond 
to so called \emph{maximal punctures} arising from the $5$d boundary 
conditions 
at the two ends of the tube. A consistency check for the resulting 
compactifications is the match of 't~Hooft anomalies of the $6$d theory on 
the Riemann surface and the $4$d theory. Since the numbers of $U(1)$ flavour 
symmetries are equal, the central charges of the two theories obtained from 
$a$-maximisation are then bound to match. 

The story developed in \cite{Gaiotto:2015usa,Razamat:2016dpl}, was later 
generalised to other $6$d SCFTs in 
\cite{Bah:2017gph,Kim:2017toz,Kim:2018bpg,Kim:2018lfo,Razamat:2018gro}. In 
\cite{Kim:2018lfo} it was realised that when compactifying the $6$d theory, 
knowledge of the resulting $5$d theory is essential for constructing domain 
walls 
and, subsequently, torus compactifications. This recipe works quite well for 
all 
of the so-called ADE conformal matter theories \cite{DelZotto:2014hpa} as well 
as for the E-string theory \cite{Kim:2017toz} and other minimal $6$d SCFTs \cite{Razamat:2018gro}. 
One common feature of all 
such compactifications studied so far is that the $6$d theory one starts with 
is 
always of generalised $A$-type. This means that only pairwise intersections 
between adjacent nodes of the $6$d tensor branch are possible and trivalent 
vertices do not appear. 

The goal of the current paper is to extend the above results to the case where 
the $6$d $\mathcal{N}=(1,0)$ theory is of generalised $D$-type. 
The simplest possibility is the case where the discriminant locus of the 
elliptic fibration of the F-theory compactification is a collection of 
$-2$-curves intersecting according to a $D$-type quiver. Compactifying such a 
theory to five dimensions yields a circular quiver with alternating $SO$ 
and $USp$ nodes \cite{Hayashi:2015vhy}. 
Starting from there, we proceed to construct $4$d 
$\mathcal{N}=1$ domain wall solutions of the resulting $5$d theory and 
successively glue them together to obtain a torus compactification from $6$d. 
In the case of $SO$ and $USp$ nodes, this process turns out to be subtle as the 
$\half$-BPS conditions of the domain wall cannot retain the full gauge 
symmetry, 
but ultimately lead to unitary subgroups of either $SO$ or $USp$ type gauge 
nodes. When gluing domain walls together by gauging flavour symmetries, one 
notices that although all cubic gauge anomalies cancel, there are still 
non-vanishing R-symmetry anomalies of type $R-G-G$. To further cancel such 
anomalies, we are forced to modify the domain wall construction of this paper 
by considering non-maximal punctures. It is then apparent that the 
$4$d theories obtained by gluing such domain walls to obtain a candidate torus 
compactification preserve a lower number of flavour symmetries than the 
corresponding $6$d parent theory. 
Moreover, carefully performing $a$-maximisation indicates that these 
candidate theories flow to free IR theories. As a consequence, potentially 
existing non-unitary operators in the theory obtained from the $6$d 
compactificiation present an open problem. 

The organisation of the paper is as follows. In Section \ref{sec:6d}, after a 
review of the 
$D$-type $6$d theory and its brane realisation, its anomaly polynomial is 
computed. 
Thereafter, a twisted compactification of the theory 
to four dimensions is performed and a subsequent integration of the anomaly 
$8$-form on the torus yields the anomaly $6$-form of the corresponding $4$d 
$\mathcal{N}=1$ SCFT. By turning on 
fluxes for flavour symmetries, the $SU(2k)$ flavour symmetry is broken to its 
Cartan subgroup $U(1)^{2k-1}$ which mixes with the $U(1)_R$-symmetry to give 
rise to a new R-symmetry in the IR. 
The resulting central charges are computed from 
$a$-maximisation. In addition, some comments on compactifications on a 2-sphere 
with punctures are given. 
In Section \ref{sec:5d} the $4$d domain wall theories are constructed by 
compactifying the $6$d theory first to five dimensions and introducing 
$\half$-BPS boundary conditions. Then the 't~Hooft anomaly coefficients are 
computed and it is shown how to choose boundary conditions which lead to the 
cancellation of all gauge and $U(1)_R$ anomalies. 
Then in Section \ref{sec:4d} all the ingredients are put 
together to compute the final $4$d quiver theory corresponding to torus 
compactification from $6$d. 
Lastly, Section \ref{sec:conclusion} provides a conclusion as well as an outline 
of open problems and possible future directions.
For convenience, Appendix \ref{app:anomalies} provides a summary of conventions 
relevant for anomaly calculations.
%
%
\section{Six dimensions}
\label{sec:6d}
This section reviews the construction of the $6$d model of $D$-type. 
Thereafter, the anomaly 8-form is derived and subsequently reduced along a 
Riemann surface. This allows the computation of $4$d central charges via 
$a$-maximisation. 

\subsection{6d theory}
The $6$d \noz of interest admits two constructions. Starting in Type IIB, one 
can consider the world-volume theory that lives on $k$ D$5$-branes transverse 
to a $D_{N+1}$ singularity. As known from the ADE quiver gauge 
theories \cite{Douglas:1996sw,Johnson:1996py}, 
the $6$d low-energy world-volume can be conveniently summarised in a 
$D_{N+1}$-Dynkin type quiver gauge theory.
Alternatively, one may employ a Type IIA brane construction  
of $N$ NS$5$ branes and $2k$ D$6$-branes in the presence of an 
$ON^0$-plane. As shown in \cite{Hanany:1999sj}, the $ON^0$-plane results in a 
$D$-type quiver gauge theory on the tensor branch of the corresponding $6$d 
\noz 
theory. 

Therefore, the $6$d \noz theory on the tensor branch includes vector and 
hypermultiplets which are coupled according to the quiver diagram
\begin{align}
	\raisebox{-.5\height}{
 	\begin{tikzpicture}
	\tikzstyle{gauge} = [circle, draw,inner sep=3pt];
	\tikzstyle{flavour} = [regular polygon,regular polygon sides=4,inner 
sep=3pt, draw];
	\node (g2) [gauge,label=below:{$\scriptscriptstyle{SU(2k)}$}] 
{};
	\node (g0) [gauge, above of=g2,label=left:{$\scriptscriptstyle{SU(k)}$}] {};
\node (g1) [gauge, left of=g2,label=below:{$\scriptscriptstyle{SU(k)}$}] 
{};
	\node (g3) [gauge, right of=g2,label=below:{$\scriptscriptstyle{SU(2k)}$}] 
{};
	\node (g4) [right of=g3] {$\ldots$};
	\node (g5) [gauge, right of=g4,label=below:{$\scriptscriptstyle{SU(2k)}$}] 
{};
	\node (f5) [flavour,right of=g5,label=below:{$\scriptscriptstyle{SU(2k)}$}] 
{};
	\draw (g0)--(g2) (g1)--(g2) (g2)--(g3) (g3)--(g4) (g4)--(g5)  (g5)--(f5);
	\draw[decoration={brace,raise=10pt},decorate,thick]
  (3.2,-0.25) -- node[below=10pt] {$\scriptstyle{N-1}$} (-0.2,-0.25);
	\end{tikzpicture}
	}\,.
	\label{eq:6d_quiver_D-type}
\end{align}
In addition, there exist $(N+1)$ tensor multiplets, one for each gauge group 
factor.
\subsection{6d anomaly polynomial}
Based on the quiver \eqref{eq:6d_quiver_D-type}, one can derive the anomaly 
polynomial via using the results of 
\cite{Ohmori:2014kda,Intriligator:2014eaa}. To arrive at the 
anomaly 8-form there are several steps to take. 
To begin with, the $6$d \noz multiplets contribute as follows:
\begin{itemize}
    \item  A hypermultiplet transforming in representation $\rho$:
 \begin{align}
  I_{8}^{\mathrm{hyper}} = \frac{1}{24} \Tr_\rho F^4 + \frac{1}{48} \Tr_\rho 
F^2\
p_1(T) + \frac{d_\rho }{5760} \left(7p_1^2(T) -4p_2(T) \right) \,,
 \end{align}
 where $d_\rho$ denotes the dimension of the representation $\rho$.
    \item A vector multiplet of gauge group $G$:
    \begin{align}
\begin{aligned}
 I_{8}^{\mathrm{vector}} = 
 &-\frac{1}{24} \left( \Tr_{\mathrm{adj}} F^4 + 6 
c_2(R) \Tr_{\mathrm{adj}} F^2 + d_G c_2(R)^2 \right) \\
&-\frac{1}{48} \left( \Tr_{\mathrm{adj}} F^2  + d_G c_2(R)\right) p_1 (T) 
-\frac{d_G }{5760} \left( 7p_1^2 (T) - 4p_2 (T) \right)
\end{aligned}
\end{align}
and $d_G$ is the dimension of $G$.
    \item A tensor multiplet:
\begin{align}
  I_{8}^{\mathrm{tensor}} = \frac{1}{24} c_2^2(R) + \frac{1}{48} c_2(R) 
p_1(T) + \frac{1 }{5760} \left(23 p_1^2(T) -116 p_2(T) \right) \,.
 \end{align}
 \end{itemize}
The notation for the appearing characteristic classes is as follows: $c_2(R)$ 
for the second Chern classes in the 
fundamental representations of the $6$d \noz $SU(2)_R$ R-symmetries; $p_1(T)$ 
and $p_2(T)$ for the first and second Pontryagin classes of the tangent bundle.
Moreover, $F_G$ denotes the field strength of the flavour symmetry 
$G=SU(2k)$; and the subscripts $\rho$, $\mathrm{f}$, $\mathrm{adj}$ of a 
trace indicates with respect to which representation $\rho$, adjoint, or 
fundamental the trace is performed.

Then one can determine the anomaly 8-form contributions from the vector 
and hypermultiplets encoded in the quiver \eqref{eq:6d_quiver_D-type} as 
well as the contributions of the $(N+1)$ tensor multiplets.
Summing all perturbative contributions of the \noz multiplets, one 
finds the following pure gauge, mixed gauge 
R-symmetry, and mixed gauge flavour anomaly terms 
\begin{align}
 I_{8}^{\mathrm{pert}} \supset -\frac{1}{8} (A_{D_{N+1}})^{ij} \Tr_f F_i^2 
\Tr_f F_j^2 
- \frac{1}{2} \rho^i \Tr_f F_i^2 c_2(R) -2 \gamma^i \Tr_f 
F_i^2 \Tr_f F_G^2 \,,
\label{eq:gauge_anomalies_6d}
\end{align}
where $i,j=1,\ldots,N+1$ labels the gauge group factors in 
\eqref{eq:6d_quiver_D-type}. The numbering of the nodes in the underlying 
$D_{N+1}$ Dynkin diagram follows the conventions of \cite[Table 
IV]{Fuchs:1997jv}, i.e.\ the spinor nodes are labelled by $N$ and $N+1$, 
respectively; while the node attached to the flavour is labelled by $i=1$.
Moreover, $A_{D_{N+1}}$ denotes the Cartan matrix of $D_{N+1}$, see \cite[Table 
VI]{Fuchs:1997jv}, and the two 
$(N+1)$-dimensional vectors $\rho$, $\gamma$ are defined as follows:
\begin{align}
 \rho =(2k,2k,\ldots,2k,k,k) \, , \qquad
 \gamma=-\frac{1}{4}(1,0,\dots,0) \,.
\end{align}
In order to cancel all pure and mixed gauge anomalies, one adds a 
Green-Schwarz term \cite{Sadov:1996zm,Kumar:2010ru,Monnier:2013kna}
\begin{align}
 I_{8}^{\mathrm{GS}} = \frac{1}{2}\Omega^{ij} I_i I_j \,,
 \label{eq:GS_term}
\end{align}
and the form of the anomalies \eqref{eq:gauge_anomalies_6d} determines 
the Green-Schwarz term almost uniquely to be 
\begin{align}
\Omega^{ij} = (A_{D_{N+1}})^{ij} \, , \qquad
 I_i = \frac{1}{2} \Tr_f F_i^2 
 + (A_{D_{N+1}}^{-1})_{ij} \left(  \rho^j c_2(R) + 2 \gamma^j \Tr_f F_G^2 
\right) 
\,. 
\end{align}
For the inverse $(A_{D_{N+1}}^{-1})_{ij}$ of the Cartan matrix the reader is 
referred to \cite[Table IV]{Fuchs:1997jv}.
Finally, adding the perturbative contribution from the supermultiplets and the 
GS-term, one arrives at the full anomaly polynomial\footnote{In view of 
\eqref{eq:6d_quiver_D-type}, the derived anomaly polynomial is valid for 
$k\geq2$. The special case $k=1$ should be considered separately.}

\begin{align}
\begin{aligned}
I_{8} = &\frac{ (8N^3-4N+1)k^2+(N+1) }{12}c_2(R)^2 
-\frac{(2N-1)k^2-(N+1)}{24} c_2(R)\,p_1(T) \\
&-\frac{k(2N-1)}{2}\Tr_f F_G^2\,c_2(R)+\frac{k}{24}\Tr_f 
F_G^2\,p_1(T)+\frac{1}{8}\left(\Tr_f F_G^2\right)^2+\frac{k}{12}\Tr_f F_G^4 \\
&+\frac{14k^2+30(N+1)}{5760} p_1(T)^2 
-\frac{8\left(k^2+15(N+1)\right)}{5760} p_2(T).
\label{eq:6d_anomlay_poly}
\end{aligned}
\end{align}
For later convenience, consider the flavour and R-symmetry bundles in more 
detail, see for instance \cite{Bah:2017gph}. Suppose the $G=SU(2k)$ flavour 
symmetry bundle splits, and the Chern roots are given by $b_i$, $i=1,\ldots,2k$ 
satisfying $\sum_{i=1}^{2k}b_i=0$. Then one finds
\begin{align}
 \Tr_f F_G^2 = - \sum_{i=1}^{2k} b_i^2
 \, , \qquad
 \Tr_f  F_G^4 = \sum_{i=1}^{2k} b_i^4 \,.
 \label{eq:Chern-root_flavour}
\end{align}
Similarly, the $SU(2)_R$ bundle splits and has Chern roots $(x,-x)$ such that 
\begin{align}
 c_2(R)=-x^2 \,.
\label{eq:Chern-root_R-sym}
\end{align}
\subsection{Anomaly polynomial after compactification}
Next, one can compute the anomaly $6$-form of a $4$d \none theory that 
originates from the compactification of the $6$d \noz theory 
\eqref{eq:6d_quiver_D-type} on a Riemann surface with fluxes via the anomaly 
$8$-form \eqref{eq:6d_anomlay_poly}. 

Generically, there are two effects to be taken into account 
when compactifiying on a genus $g$ Riemann surface $C_g$ with fluxes. Firstly, 
to preserve \none supersymmetry in $4$d one must perform a twist. Roughly, the 
$6$d Lorentz group decomposes into the $4$d Lorentz group times an $SO(2)$ 
acting on $C_g$. Breaking the  $6$d $SU(2)_R$ symmetry to a maximal 
torus $U(1)_R$, which is the natural candidate for 
the $4$d R-symmetry, one then twists it with the $SO(2)$ to ensure \none 
supersymmetry in $4$d.
Consequently, the Pontryagin classes decompose as \cite{Razamat:2016dpl}
\begin{align}
 p_1(T)= t^2 +p_1(T') \,, \qquad
 p_2(T)=t^2 p_1(T') +p_2(T')
\end{align}
into $4$d Pontryagin classes 
of the tangent bundle, $p_1(T')$ and $p_2(T')$,  and the first Chern class of 
the Riemann surface, $t$.
For the R-symmetry, the twisted compactifiaction leads to a mixing between 
the spin connection $t$ on $C_g$ and $c_1(R')$, the first Chern class of 
the $U(1)_R$ bundle, such that the Chern root becomes 
\begin{align}
 x = c_1(R') -\frac{1}{2} t \,. 
 \label{eq:relation_Chern-roots_R}
\end{align}
Secondly, the flavour symmetry fluxes break the $SU(2k)$ symmetry to 
its torus too, i.e.
\begin{align}
G=SU(2k)\longrightarrow U(1)^{2k-1}\,.
\end{align}
Denote by $z_i$, $i=1,\ldots,2k$ the fluxes for the Cartan generators 
of $\mathfrak{u}(1)_{b_i}$ of the $6$d flavour symmetry and suppose 
$c_1(\beta_i)$ are the first Chern classes of line bundles in $4$d.
Again, the first Chern class of $C_g$ mixes with flavour symmetries and the  
Chern roots are related via \cite{Bah:2017gph}
\begin{align}
 b_i = N c_1(\beta_i) - z_i \frac{t}{2g-2 } \,.
 \label{eq:relation_Chern-roots_flavour}
\end{align}
The constraint $\sum_{i=1}^{2k} b_i =0$  then implies 
\begin{align}
 \sum_{i=1}^{2k} c_1(\beta_i) =0 \, , \qquad
 \sum_{i=1}^{2k} z_i =0 \,.
 \label{eq:constraints}
\end{align}
Note that the Gauss-Bonnet theorem $\int_{C_g} t = 2-2g$ leads to
\begin{align}
 \int_{C_g} b_i = z_i\,,
\end{align}
which is a measure for the flux on the torus.

Now, to compute the resulting $4$d anomaly $6$-form one starts from the anomaly 
$8$-form \eqref{eq:6d_anomlay_poly}, inserts the splitting of flavour bundles 
\eqref{eq:Chern-root_flavour} and R-symmetry bundles 
\eqref{eq:Chern-root_R-sym}, translates the $6$d objects via 
\eqref{eq:relation_Chern-roots_R}, \eqref{eq:relation_Chern-roots_flavour} into 
$4$d quantities, and lastly integrates 
over the Riemann surface $C_g$. After careful evaluation, one finds
\begin{align}
\label{eq:4d_anomaly_full}
\begin{aligned}
 I_6= 
&\frac{(g-1)}{3}   \left(k^2 \left(8 N^3-4 N+1\right)+N+1\right) c_1(R')^3 \\
&- kN (2 N-1)  \sum _{i=1}^{2 k} z_i \ c_1(\beta_i)  c_1(R')^2 
 -(g-1)  k (2 N-1 ) N^2 \sum_{i=1}^{2 k} c_1(\beta_i)^2  c_1(R')\\
&+\frac{(g-1)}{12}\left(k^2 (2 N-1)-(N+1)\right) \ c_1(R') p_1(T') 
-\frac{k N}{12} \sum _{i=1}^{2 k}z_i \  c_1(\beta_i)  p_1(T')  \\
&+\frac{N^3}{12} \left(
3\sum _{i,j=1}^{2 k} \left(z_i \ c_1(\beta_i) 
c_1(\beta_j)^2 + z_j\ c_1(\beta_i)^2 c_1(\beta_j) \right)
+4 k \sum _{i=1}^{2 k} z_i\ c_1(\beta_i)^3 \right)
\end{aligned}
\end{align}
which still needs to be supplemented by the constraints \eqref{eq:constraints}. 
\paragraph{Example: 2-torus.}
Specialising the result to the torus $T^2$ requires a comment as $g=1$: the 
singular-looking flux contribution $\tfrac{t}{2g-2}$ is then understood as 
2-form such that the integral over the Riemann surface is $-1$, 
cf.\ \cite[Footnote 2]{Bah:2017gph}. With this in mind, the anomaly 6-form 
reads as
\begin{align}
\label{eq:4d_anomaly_torus}
\begin{aligned}
 I_6\big|_{T^2}= 
&- kN (2 N-1)  \sum _{i=1}^{2 k} z_i \ c_1(\beta_i)  c_1(R')^2 
-\frac{k N}{12} \sum _{i=1}^{2 k}z_i \  c_1(\beta_i)  p_1(T')  \\
&+\frac{N^3}{12} \left(
3\sum _{i,j=1}^{2 k} \left(z_i \ c_1(\beta_i) 
c_1(\beta_j)^2 + z_j\ c_1(\beta_i)^2 c_1(\beta_j) \right)
+4 k \sum _{i=1}^{2 k} z_i\ c_1(\beta_i)^3 \right) \,.
\end{aligned}
\end{align}
As a remark, the anomaly $6$-form \eqref{eq:4d_anomaly_torus} clearly shows 
that 
the $4$d gravity anomalies $\Tr( U(1)_R) $, $\Tr( U(1)_R^3) $  vanish in the 
UV. 

\paragraph{Example: 2-sphere with $s$ punctures.}
Considering a 2-sphere with $s$ punctures can be achieved by replacing $g \to 
g +\frac{1}{2}s$, then imposing $g=0$ yields
\begin{align}
\label{eq:4d_anomaly_punct-sphere}
\begin{aligned}
 I_6\big|_{S^2_s}= 
&\frac{(s-2)}{6}   \left(k^2 \left(8 N^3-4 N+1\right)+N+1\right) 
c_1(R')^3 \\
&- kN (2 N-1)  \sum _{i=1}^{2 k} z_i \ c_1(\beta_i)  c_1(R')^2 
 -\frac{(s-2)  k (2 N-1 ) N^2}{2} \sum_{i=1}^{2 k} c_1(\beta_i)^2  
c_1(R')\\
&+\frac{(s-2)}{24}\left(k^2 (2 N-1)-(N+1)\right) \ c_1(R') p_1(T') 
-\frac{k N}{12} \sum _{i=1}^{2 k}z_i \  c_1(\beta_i)  p_1(T')  \\
&+\frac{N^3}{12} \left(
3\sum _{i,j=1}^{2 k} \left(z_i \ c_1(\beta_i) 
c_1(\beta_j)^2 + z_j\ c_1(\beta_i)^2 c_1(\beta_j) \right)
+4 k \sum _{i=1}^{2 k} z_i\ c_1(\beta_i)^3 \right)
\,.
\end{aligned}
\end{align}
In addition, one has to take the contributions from the punctures into account 
too, which are essentially constant shifts to the IR $a$-central charge, see 
for 
instance \cite{Razamat:2018gro}. This additional contribution is discussed in 
Section \ref{sec:non-max_bc} and becomes relevant in Section 
\ref{sec:quiver_2-sphere+puncture}.
%
%
\subsection{\texorpdfstring{$a$}{a}-maximisation}
\label{sec:a-max_6d}
Next, consider the $a$-maximisation \cite{Intriligator:2003jj} of the $4$d 
\none 
theory with anomaly polynomial \eqref{eq:4d_anomaly_full}, assuming that the 
constraints \eqref{eq:constraints} are imposed. 
The trial $R$-charge is a linear combination of the UV $U(1)_R$ $R$-charge and 
the 
different $U(1)_{b_i}$ from the maximal torus of the $6$d flavour symmetry, 
i.e.
\begin{align}
U(1)_R^{\mathrm{trial}} = U(1)_R + \sum_{i=1}^{2k-1} x_i \ U(1)_{b_i} 
\quad \text{s.t.} \quad 
R_\mathrm{trial} = R_{UV} + \sum_{i=1}^{2k-1} x_i \beta_i 
\, , \quad x_i \in \mathbb{R}  \,.
\end{align}
Then the trial $a$-central charge becomes
\begin{align}
 a_{\mathrm{trial}} &= \frac{3}{32} \left( 3 \Tr R_\mathrm{trial}^3 - \Tr 
R_\mathrm{trial} \right) \notag \\
&=\frac{3}{32} \Big( 3 \Tr( R_{UV}^3) 
+ 9 \sum_{i} x_i \Tr(R_{UV}^2 \beta_i )
+ 9 \sum_{i,j} x_i x_j \Tr(R_{UV} \beta_i \beta_j) 
 \notag \\
& \qquad + 3 \sum_{i,j,k} x_i x_j x_k \Tr( \beta_i \beta_j \beta_k)
-\Tr( R_{UV} )- \sum_{i} x_i \Tr( \beta_i )
\Big) \,.
\end{align}
The trace coefficients can be read off from \eqref{eq:4d_anomaly_torus}; see 
examples below.
Next, the $a$-maximisation procedure requires to solve
\begin{align}
 \frac{\partial a_{\mathrm{trial}} }{\partial x_i} =0 \, , \qquad \forall i = 
1,\ldots, 2k-1\,. 
\label{eq:a_max}
\end{align}
However, due to the large number of equations as well as the equally large 
number of free fluxes $z_i$, the analytic evaluation is cumbersome. To gain 
some understanding, one may resort to examples with low value of $k$ and equal 
fluxes $z_i=z$ for all $i=1,\ldots,2k-1$.

For the 2-torus, one finds the following contributions to the trial central 
charge:
\begin{subequations}
\label{eq:trace_terms_torus}
\begin{align}
\Tr( R_{UV} ) &=-24 \cdot \left[ I_{6}\big|_{T^2}\right]_{c_1(R') p_1(T')} = 0 
\,, \\
\Tr( R_{UV}^3 ) &=6 \cdot \left[ I_{6}\big|_{T^2}\right]_{c_1(R')^3} = 0 \,, \\
\Tr(\beta_i) &= -24 \cdot \left[ 
I_{6}\big|_{T^2}\right]_{c_1(\beta_i) p_1(T')} 
\,, \\
\Tr( \beta_i \beta_j \beta_k) &= d_{ijk} \cdot \left[ 
I_{6}\big|_{T^2}\right]_{c_1(\beta_i) c_1(\beta_j) c_1(\beta_k) } \,, \\
\Tr( R_{UV}^2  \beta_i ) &= 2 \cdot \left[ 
I_{6}\big|_{T^2}\right]_{c_1(R')^2 c_1(\beta_i) } \,, \\
\Tr( R_{UV}  \beta_i \beta_j) &= d_{ij} \cdot \left[ 
I_{6}\big|_{T^2}\right]_{c_1(R') c_1(\beta_i) c_1(\beta_j)}  \,,
\end{align}
\end{subequations}
where $\left[ I_{6}\big|_{T^2}\right]_{X} $ denotes the coefficient of 
the combination of characteristic classes $X$ in the anomaly $6$-form 
\eqref{eq:4d_anomaly_torus}. Moreover, $d_{i_1 i_2 \ldots i_n}$ equals $m!$ 
with $m$ being the number of equal indices in $i_1 i_2 \ldots i_n$.
\paragraph{Example $k=2$.}
For $k=2$ and equal fluxes, the analytic solution to \eqref{eq:a_max} is found 
to be
\begin{align}
 x_i = -\frac{\sqrt{9N -4}}{6 \sqrt{2}N} \, ,\; i=1,2,3
 \quad \Rightarrow \quad 
 a = \frac{z }{\sqrt{2}} \left(9N -4\right)^{\frac{3}{2}} \,.
 \label{eq:ex_T2_k=2}
\end{align}
\paragraph{Example $k=3$.}
For $k=3$ and equal fluxes, the analytic solution to \eqref{eq:a_max} is found 
to be
\begin{align}
 x_i = -\frac{\sqrt{9N -4}}{9 \sqrt{2}N} \,,\; i=1,\ldots,5
 \quad \Rightarrow \quad 
 a = \frac{5z }{2 \sqrt{2}} \left(9N -4\right)^{\frac{3}{2}} \,.
 \label{eq:ex_T2_k=3}
\end{align}
\paragraph{Example $k=4$.}
For $k=4$ and equal fluxes, the analytic solution to \eqref{eq:a_max} is found 
to be
\begin{align}
 x_i = -\frac{\sqrt{9N -4}}{12 \sqrt{2}N} \,,\; i=1,\ldots,7
 \quad \Rightarrow \quad 
 a =  \frac{7\sqrt{2} z}{3 } \left(9N -4\right)^{\frac{3}{2}} \,.
 \label{eq:ex_T2_k=4}
\end{align}
%
%
%
%
\section{Five dimensions}
\label{sec:5d}
The $6$d theory \eqref{eq:6d_quiver_D-type} can be compactified on $S^1$. The 
resulting $5$d theory has a low-energy description in terms of a $5$d \none 
affine $A_{2k-1}$ quiver gauge theory with alternating $SO(2k{+}2)$ and 
$USp(2k{-}2)$ gauge nodes \cite{Hayashi:2015vhy}, i.e.
\begin{align}
	\raisebox{-.5\height}{
 	\begin{tikzpicture}
	\tikzstyle{gauge} = [circle, draw,inner sep=3pt];
	\tikzstyle{flavour} = [regular polygon,regular polygon sides=4,inner 
sep=3pt, draw];
	\node (g1) [gauge,label=below left:{$\scriptscriptstyle{SO(2N{+}2)}$}] 
{};
	\node (g2) [gauge, right 
of=g1,label=below:{$\scriptscriptstyle{USp(2N{-}2)}$}] 
{};
	\node (g3) [right of=g2] {$\ldots$};
	\node (g4) [gauge, right 
of=g3,label=below:{$\scriptscriptstyle{SO(2N{+}2)}$}] 
{};
	\node (g5) [gauge,right 
of=g4,label=below right:{$\scriptscriptstyle{USp(2N{-}2)}$}] 
{};
	\node (g6) [gauge, above 
of=g5,label=above right:{$\scriptscriptstyle{SO(2N{+}2)}$}] 
{};
	\node (g7) [gauge,left 
of=g6,label=above:{$\scriptscriptstyle{USp(2N{-}2)}$}] 
{};
	\node (g8) [left of=g7] {$\ldots$};
	\node (g9) [gauge, left 
of=g8,label=above:{$\scriptscriptstyle{SO(2N{+}2)}$}] 
{};
	\node (g10) [gauge,left 
of=g9,label=above left:{$\scriptscriptstyle{USp(2N{-}2)}$}] 
{};
	\draw (g1)--(g2) (g2)--(g3) (g3)--(g4) (g4)--(g5) (g5)--(g6) (g6)--(g7) 
(g7)--(g8) (g8)--(g9) (g9)--(g10) (g10)--(g1);
	\draw[decoration={brace,raise=10pt},decorate,thick]
  (4.5,1.05) -- node[right=10pt] {$\substack{ \scriptstyle{k \ \times \ 
SO(2N{+}2) \text{ nodes}\, , } \\ \scriptstyle{k \ \times \ USp(2N{-}2) \text{ 
nodes} } }$} (4.5,-0.05);
	\end{tikzpicture}
	}\,.
	\label{eq:5d_quiver_affine_A}
\end{align}
This $5$d theory will be the basis for the construction of flux domain wall 
theories in the spirit of \cite{Gaiotto:2015una}, see also 
\cite{Kim:2017toz,Kim:2018bpg,Chan:2000qc}.
%
%
\subsection{\texorpdfstring{$\tfrac{1}{2}$}{half} BPS boundary conditions}
\label{sec:5d_half_BPS_bc}
Similar to the approach taken in \cite{Kim:2018lfo}, one considers the boundary 
conditions for vector and hypermultiplets that need to be imposed in order to 
define a domain wall. 
\paragraph{Preserve symplectic gauge group.}
To begin with, focus on the $USp(2N{-}2)$ gauge nodes and 
impose Neumann boundary conditions on the gauge field, i.e.\
\begin{align}
A_4^{USp(2N{-}2)}\vert_{x^4=0}=0= \partial_4 
A^{USp(2N{-}2)}_\mu\vert_{x^4=0}\,, 
\for \mu=0,1,2,3\,,
\label{eq:USp1}
\end{align}
which preserve the full gauge group. On the other hand,  Dirichlet 
boundary conditions are imposed on the adjoint chiral $\Phi^{USp(2N{-}2)}$ in 
the $5$d $\mc N=1$ gauge multiplet, i.e.
\begin{align}
\Phi^{USp(2N-2)}\vert_{x^4=0}=0 \,.
\label{eq:USp4}
\end{align}
Next, the hypermultiplets connecting a 
$USp(2N{-}2)$ and a $SO(2N{+}2)$ gauge node transform in the fundamental 
representation of $USp(2N{-}2)$ and in the vector representation of 
$SO(2N{+}2)$. Hence, viewed from the $USp(2N{-}2)$ gauge node there are $(N+1)$ 
copies of fundamental hypers, i.e.
\begin{align}
H_i=(X_i,\,Y_i^\dagger)\,, \for i=1,2,\dots N+1\,.
\end{align}
Unlike the case of bifundamental hypermultiplets of unitary gauge groups, the 
two \none chiral multiplets are not in inequivalent representations. To see 
this, recall that fundamental representation of $USp(2N{-}2)$ is pseudoreal; 
hence, the flavour symmetry group for $(N+1)$ copies enhances from $SU(N{+}1)$ 
to 
$SO(2N{+}2)$.
Turning to boundary conditions for $H_i$, one has the following two options for 
each $i=1,2,\ldots, N+1$ that preserve half the supersymmetries:
\begin{align}
+) \quad \partial_4 X_i\vert_{x^4=0}=Y_i\vert_{x^4=0}=0\,,\ 
\quad \text{or} \quad    
-)\quad
X_i\vert_{x^4=0}=\partial_4 Y_i\vert_{x^4=0}=0\,.
\label{eq:USp2}
\end{align} 
Denote the $(N+1)$-dimensional vector of boundary conditions by $\sigma$ 
with components $\sigma_i = \pm$. 
Naively, one would conclude that there are $2^{N+1}$ choices for all $N+1$ 
hypermultiplets. However, since $X_i$ and $Y_i^\dagger$ transform in equivalent 
representations of $USp(2N{-}2)$, any of these choices leads to $(N+1)$ 
chiral multiplets in the fundamental $USp(2N{-}2)$ representation. 
Consequently, 
the $(N+1)$ chiral multiplets admit at most a $SU(N{+}1)\subset SO(2N{+}2)$ 
flavour 
symmetry. The important question is, whether the choice $\sigma$ of 
boundary conditions implies that the chirals transform in the fundamental or 
anti-fundamental representation of $SU(N{+}1)$. Consider the two extreme cases 
$\sigma_\pm \equiv (\pm,\ldots,\pm)$, meaning either all $X_i$ survive 
for $\sigma_+$ or all $Y_i$ survive for $\sigma_-$.  Then the origin of 
the enhanced flavour symmetry implies that the $X_i$ furnish the fundamental 
and the $Y_i$ the anti-fundamental representation of $SU(N{+}1)$.
The choice of a generic boundary condition differs from the extreme case only 
by a different embedding $SU(N{+}1) \hookrightarrow SO(2N{+}2)$. In particular, 
if 
$\sigma$ is a fixed choice, then there exists the ''opposite`` choice 
$-\sigma$, where all signs are reverted, such that $-\sigma$ 
corresponds to the same embedding, but the surviving $(N+1)$ chirals transform 
in the conjugate $SU(N{+}1)$ representation compared to the ones from 
$\sigma$. Consequently, it is sufficient to consider the extreme cases 
$\sigma_\pm$. 

Turning to the $SO(2N{+}2)$ gauge nodes, it is suggestive to impose Neumann 
boundary conditions on the $SU(N{+}1)$ subgroup resulting from the chiral 
matter 
fields and set the remaining gauge field components to zero, i.e.
\begin{align}
 A_{4}^{SO(2N{+}2)} \vert_{x^4=0} 
 =
\partial_4 A^{SO(2N{+}2)}_{\mu} \vert_{x^4=0} 
=A_{\mu}^{SO(2N{+}2)\backslash SU(N{+}1)} \vert_{x^4=0}=0 
\for \mu=0,1,2,3 \,.
\label{eq:USp3}
\end{align}
Meanwhile, Dirichlet boundary conditions are imposed on the full adjoint chiral 
of the $5$d $\mc N=1$ $SO(2N+2)$ gauge multiplet, i.e.
\begin{align}
\Phi^{SO(2N+2)}\vert_{x^4=0}=0 \,.
\label{eq:USp5}
\end{align}

Therefore, the boundary conditions defined in \eqref{eq:USp1}, \eqref{eq:USp4}, \eqref{eq:USp2}, \eqref{eq:USp3} and \eqref{eq:USp5} specify one chamber of the $5$d theory on the interface. 
For each of the $2k$ hypermultiplets in \eqref{eq:5d_quiver_affine_A} one may 
choose boundary conditions $\sigma_\pm$, such that the theory is determined 
by a $(2k)$-dimensional vector $\mathbb{B}$.  Here, the convention is that 
$\sigma_+$ turns the hypermultiplets into a chiral in the fundamental 
representation of $SU(N{+}1)$, while $\sigma_-$ corresponds to the 
anti-fundamental. To exemplify a few cases, one may consider
\begin{subequations}
\begin{align}
\mathbb{B}= (\ldots,+,+,-,-,+,\ldots)
\qquad
	\raisebox{-.5\height}{
 	\begin{tikzpicture}
	\tikzstyle{gauge} = [circle, draw,inner sep=3pt];
	\tikzstyle{flavour} = [regular polygon,regular polygon sides=4,inner 
sep=3pt, draw];
\node (g1) [gauge,label=below:{$\scriptscriptstyle{USp(2N{-}2)}$}] 
{};
\node (g0) [left of=g1] {$\ldots$};
\node (g2) [gauge, right of=g1,label=above:{$\scriptscriptstyle{SU(N{+}1)}$}] 
{};
\node (g3) [gauge,right of=g2,label=below:{$\scriptscriptstyle{USp(2N{-}2)}$}] 
{};
\node (g4) [gauge, right of=g3,label=above:{$\scriptscriptstyle{SU(N{+}1)}$}] 
{};
\node (g5) [right of=g4] {$\ldots$};
\begin{scope}[decoration={markings,mark =at position 0.5 with 
{\arrow{stealth}}}]
    \draw[postaction={decorate}] (g0)--(g1);
    \draw[postaction={decorate}] (g2) -- (g1);
    \draw[postaction={decorate}] (g3)--(g2);
    \draw[postaction={decorate}] (g3)--(g4);
    \draw[postaction={decorate}] (g4) -- (g5);
\end{scope}
\end{tikzpicture}
	}\,, \\
\mathbb{B}= (\ldots,-,-,-,+,+,\ldots)
\qquad
	\raisebox{-.5\height}{
 	\begin{tikzpicture}
	\tikzstyle{gauge} = [circle, draw,inner sep=3pt];
	\tikzstyle{flavour} = [regular polygon,regular polygon sides=4,inner 
sep=3pt, draw];
\node (g1) [gauge,label=below:{$\scriptscriptstyle{USp(2N{-}2)}$}] 
{};
\node (g0) [left of=g1] {$\ldots$};
\node (g2) [gauge, right of=g1,label=above:{$\scriptscriptstyle{SU(N{+}1)}$}] 
{};
\node (g3) [gauge,right of=g2,label=below:{$\scriptscriptstyle{USp(2N{-}2)}$}] 
{};
\node (g4) [gauge, right of=g3,label=above:{$\scriptscriptstyle{SU(N{+}1)}$}] 
{};
\node (g5) [right of=g4] {$\ldots$};
\begin{scope}[decoration={markings,mark =at position 0.5 with 
{\arrow{stealth}}}]
    \draw[postaction={decorate}] (g1)--(g0);
    \draw[postaction={decorate}] (g1)--(g2);
    \draw[postaction={decorate}] (g3)--(g2);
    \draw[postaction={decorate}] (g4)--(g3);
    \draw[postaction={decorate}] (g4)--(g5);
\end{scope}
\end{tikzpicture}
	}\,.
\end{align}
\label{eq:USp2max}
\end{subequations}
\paragraph{Preserve orthogonal gauge group.}
On the other hand, one can equally well impose Neumann boundary conditions on 
the $SO(2N{+}2)$  gauge nodes to preserve the entire gauge group, i.e.
\begin{align}
 A_4^{SO(2N{+}2)}\vert_{x^4=0}=\partial_4 A^{SO(2N{+}2)}_\mu\vert_{x^4=0}=0\,, 
\for 
\mu=0,1,2,3 \,,
\label{eq:SO1}
\end{align}
and similarly Dirichlet boundary conditions on the adjoint chiral 
$\Phi^{SO(2N+2)}$ in the $5$d $\mc N=1$ $SO(2N+2)$ gauge multiplet,
\begin{align}
\Phi^{SO(2N+2)}\vert_{x^4=0}=0\,.
\label{eq:SO4}
\end{align} 
Viewed from the $SO(2N{+}2)$ gauge group, the hypermultiplets 
\begin{align}
 H^\prime_{a}=(X^\prime_a,\,Y_a^{\prime\dagger}) \for a=1,\dots,N-1
\end{align}
between this gauge group and the adjacent gauge node are understood as $(N-1)$ 
copies of fundamental $SO(2N{+}2)$ hypermultiplets. Again, the two \none 
chirals 
in each $H^\prime_a$ are in equivalent $SO(2N{+}2)$ representations, which is 
the 
reason for the flavour symmetry enhancement $SU(N{-}1) \to USp(2N{-}2)$. 
By assigning $\frac{1}{2}$ BPS boundary conditions to the hypermultiplets, one 
has two 
choices 
\begin{align}
 +') \quad \partial_4 
X^\prime_a\vert_{x^4=0}=Y^\prime_a\vert_{x^4=0}=0\,,\ 
\quad \text{or} \quad
-') \quad 
X^\prime_a\vert_{x^4=0}=\partial_4 Y^\prime_a\vert_{x^4=0}=0 \,,
\label{eq:SO2}
\end{align}
for each  $a=1,\dots,N-1$. Nevertheless, all of these choices result in $(N-1)$ 
chiral multiplets in the fundamental $SO(2N{+}2)$  representation. The flavour 
symmetry of these $(N-1)$ chirals is at most $SU(N{-}1)\subset USp(2N{-}2)$, 
but 
one has to determine whether they transform in the fundamental or 
anti-fundamental representation. Analogous to the above arguments, one could 
summarise the choices in \eqref{eq:SO2} in a $(N-1)$-dimensional vector 
$\sigma'$, 
with components $\pm$. Again, the extreme cases $\sigma'_\pm \equiv 
(\pm,\ldots,\pm)$ indicate that choosing $X'_a$ only leads to chirals in the 
fundamental of $SU(N{-}1)$, while the $Y'_a$ choice yields anti-fundamental 
chirals. Any other choice of $\sigma'$ characterises a different embedding 
$SU(N{-}1)\hookrightarrow USp(2N{-}2)$, but effectively reduces to chirals 
transforming in the fundamental or anti-fundamental of $SU(N{-}1)$ for either 
$\sigma'$ or the opposite choice $-\sigma'$. Thus, the two extreme cases are 
sufficient.
Next,
one should impose boundary conditions on the $USp(2N{-}2)$ gauge multiplet 
that are 
compatible with the choice of an $SU(N{-}1)$ subgroup. In detail,
\begin{equation}
\label{eq:SO3}
\begin{aligned}
 A_4^{USp(2N{-}2)}\vert_{x^4=0}
 =\partial_4 A^{USp(2N{-}2)}_\mu\vert_{x^4=0}
 =A_\mu^{USp(2N{-}2)\backslash SU(N{-}1)} \vert_{x^4=0}&=0 \for \mu=0,1,2,3 \,, 
 \\
 {\rm and}\ \ \Phi^{USp(2N{-}2)}\vert_{x^4=0}&=0 \,.
\end{aligned}
\end{equation}
The boundary conditions of \eqref{eq:SO1}, \eqref{eq:SO4}, \eqref{eq:SO2}, and \eqref{eq:SO3} 
determine another type of chamber of the $5$d theory on the interface which is 
specified by an $(2k)$-dimensional vector $\mathbb{B}'$. The convention is as 
above, if $\sigma'_+$ then the surviving chiral multiplets are $SU(N{-}1)$ 
fundamentals, while anti-fundamentals for $\sigma'_-$. The resulting theory can 
be illustrated in a few examples:
\begin{subequations}
\begin{align}
\mathbb{B}'= (\ldots,+,+,-,-,+,\ldots)
\qquad
	\raisebox{-.5\height}{
 	\begin{tikzpicture}
	\tikzstyle{gauge} = [circle, draw,inner sep=3pt];
	\tikzstyle{flavour} = [regular polygon,regular polygon sides=4,inner 
sep=3pt, draw];
\node (g1) [gauge,label=below:{$\scriptscriptstyle{SO(2N{+}2)}$}] 
{};
\node (g0) [left of=g1] {$\ldots$};
\node (g2) [gauge, right of=g1,label=above:{$\scriptscriptstyle{SU(N{-}1)}$}] 
{};
\node (g3) [gauge,right of=g2,label=below:{$\scriptscriptstyle{SO(2N{+}2)}$}] 
{};
\node (g4) [gauge, right of=g3,label=above:{$\scriptscriptstyle{SU(N{-}1)}$}] 
{};
\node (g5) [right of=g4] {$\ldots$};
\begin{scope}[decoration={markings,mark =at position 0.5 with 
{\arrow{stealth}}}]
    \draw[postaction={decorate}] (g0)--(g1);
    \draw[postaction={decorate}] (g2)--(g1);
    \draw[postaction={decorate}] (g3)--(g2);
    \draw[postaction={decorate}] (g3)--(g4);
    \draw[postaction={decorate}] (g4)--(g5);
\end{scope}
	\end{tikzpicture}
	}\,, \\
\mathbb{B}'= (\ldots,-,-,-,+,+,\ldots)
\qquad
	\raisebox{-.5\height}{
 	\begin{tikzpicture}
	\tikzstyle{gauge} = [circle, draw,inner sep=3pt];
	\tikzstyle{flavour} = [regular polygon,regular polygon sides=4,inner 
sep=3pt, draw];
\node (g1) [gauge,label=below:{$\scriptscriptstyle{SO(2N{+}2)}$}] 
{};
\node (g0) [left of=g1] {$\ldots$};
\node (g2) [gauge, right of=g1,label=above:{$\scriptscriptstyle{SU(N{-}1)}$}] 
{};
\node (g3) [gauge,right of=g2,label=below:{$\scriptscriptstyle{SO(2N{+}2)}$}] 
{};
\node (g4) [gauge, right of=g3,label=above:{$\scriptscriptstyle{SU(N{-}1)}$}] 
{};
\node (g5) [right of=g4] {$\ldots$};
\begin{scope}[decoration={markings,mark =at position 0.5 with 
{\arrow{stealth}}}]
    \draw[postaction={decorate}] (g1)--(g0);
    \draw[postaction={decorate}] (g1)--(g2);
    \draw[postaction={decorate}] (g3)--(g2);
    \draw[postaction={decorate}] (g4)--(g3);
    \draw[postaction={decorate}] (g4)--(g5);
\end{scope}
	\end{tikzpicture}
	}\,.
\end{align}
\end{subequations}
%
%
\subsection{Flux domain walls}
\label{sec:5d_flux_domain_wall}
With the preparation from Section \ref{sec:5d_half_BPS_bc}, one can construct 
a flux domain wall as the interface theories between two $5$d chambers. From 
the perspective of the original $6$d theories, the domain wall theories can be 
regarded as compactifications of the $6$d theories on a 
tube or, say, a sphere with two punctures. These punctures are then 
associated to the gauge groups of the $5$d quiver theory. If one removes the 
gauge multiplets, they will be treated as the corresponding non-Abelian global 
symmetries, in addition to the $6$d global symmetries $SU(2k)$. More 
concretely, in the cases introduced in Section \ref{sec:5d_half_BPS_bc}, there 
exist two types of puncture symmetries:  $USp(2N{-}2)$-$SU(N{+}1)$ and 
$SO(2N{+}2)$-$SU(N{-}1)$. Consequently, three types of fundamental domain 
walls arise:
\begin{subequations}
\label{eq:DW_sketch}
\begin{align}
\raisebox{-.2\height}{
 	\begin{tikzpicture}
 	\tikzset{node distance = 1.5cm}
	\tikzstyle{gauge} = [circle, draw,inner sep=3pt];
	\tikzstyle{flavour} = [regular polygon,regular polygon sides=4,inner 
sep=3pt, draw];
%
\node (g1) [flavour, label= left:{$\scriptscriptstyle{SU(N{+}1)}$}]{};
\node (g2) [flavour,below of=g1,label= 
left:{$\scriptscriptstyle{USp(2N{-}2)}$}] {};
%
%
 	\tikzset{node distance = 0.75cm}
    \node (Lup) [above of= g1] {};
    \node (Ldown) [below of= g2] {};
\draw (g1)--(g2) (g1)--(Lup) (g2)--(Ldown);
	\end{tikzpicture}
	}
\begin{tikzpicture}[rotate=-90,scale=0.5]
\fill[top color=gray!50!black,bottom color=gray!10,middle 
color=gray,shading=axis,opacity=0.10] (0,0) circle (2cm and 0.5cm);
\fill[left color=gray!50!black,right color=gray!50!black,middle 
color=gray!50,shading=axis,opacity=0.10] (2,0) -- (2,6) arc (360:180:2cm and 
0.5cm) -- (-2,0) arc (180:360:2cm and 0.5cm);
\fill[top color=gray!90!,bottom color=gray!2,middle 
color=gray!30,shading=axis,opacity=0.10] (0,6) circle (2cm and 0.5cm);
\draw (-2,6) -- (-2,0) arc (180:360:2cm and 0.5cm) -- (2,6) ++ (-2,0) circle 
(2cm and 0.5cm);
\draw[densely dashed] (-2,0) arc (180:0:2cm and 0.5cm);
\end{tikzpicture}
\raisebox{-.2\height}{
 	\begin{tikzpicture}
 	\tikzset{node distance = 1.5cm}
	\tikzstyle{gauge} = [circle, draw,inner sep=3pt];
	\tikzstyle{flavour} = [regular polygon,regular polygon sides=4,inner 
sep=3pt, draw];
%
\node (g1) [flavour, label= right:{$\scriptscriptstyle{USp(2N{-}2)}$}]{};
\node (g2) [flavour,below of=g1,label= 
right:{$\scriptscriptstyle{SU(N{+}1)}$}] {};
%
%
 	\tikzset{node distance = 0.75cm}
    \node (Lup) [above of= g1] {};
    \node (Ldown) [below of= g2] {};
\draw (g1)--(g2) (g1)--(Lup) (g2)--(Ldown);
	\end{tikzpicture}
	}
	\label{eq:Sp-SU-DW}
	\\
\raisebox{-.2\height}{
 	\begin{tikzpicture}
 	\tikzset{node distance = 1.5cm}
	\tikzstyle{gauge} = [circle, draw,inner sep=3pt];
	\tikzstyle{flavour} = [regular polygon,regular polygon sides=4,inner 
sep=3pt, draw];
%
\node (g1) [flavour, label= left:{$\scriptscriptstyle{SU(N{+}1)}$}]{};
\node (g2) [flavour,below of=g1,label= 
left:{$\scriptscriptstyle{SO(2N{+}2)}$}] {};
%
%
 	\tikzset{node distance = 0.75cm}
    \node (Lup) [above of= g1] {};
    \node (Ldown) [below of= g2] {};
\draw (g1)--(g2) (g1)--(Lup) (g2)--(Ldown);
	\end{tikzpicture}
	}
\begin{tikzpicture}[rotate=-90,scale=0.5]
\fill[top color=gray!50!black,bottom color=gray!10,middle 
color=gray,shading=axis,opacity=0.10] (0,0) circle (2cm and 0.5cm);
\fill[left color=gray!50!black,right color=gray!50!black,middle 
color=gray!50,shading=axis,opacity=0.10] (2,0) -- (2,6) arc (360:180:2cm and 
0.5cm) -- (-2,0) arc (180:360:2cm and 0.5cm);
\fill[top color=gray!90!,bottom color=gray!2,middle 
color=gray!30,shading=axis,opacity=0.10] (0,6) circle (2cm and 0.5cm);
\draw (-2,6) -- (-2,0) arc (180:360:2cm and 0.5cm) -- (2,6) ++ (-2,0) circle 
(2cm and 0.5cm);
\draw[densely dashed] (-2,0) arc (180:0:2cm and 0.5cm);
\end{tikzpicture}
\raisebox{-.2\height}{
 	\begin{tikzpicture}
 	\tikzset{node distance = 1.5cm}
	\tikzstyle{gauge} = [circle, draw,inner sep=3pt];
	\tikzstyle{flavour} = [regular polygon,regular polygon sides=4,inner 
sep=3pt, draw];
%
\node (g1) [flavour, label= right:{$\scriptscriptstyle{SO(2N{+}2)}$}]{};
\node (g2) [flavour,below of=g1,label= 
right:{$\scriptscriptstyle{SU(N{+}1)}$}] {};
%
%
 	\tikzset{node distance = 0.75cm}
    \node (Lup) [above of= g1] {};
    \node (Ldown) [below of= g2] {};
\draw (g1)--(g2) (g1)--(Lup) (g2)--(Ldown);
	\end{tikzpicture}
	}
		\label{eq:SO-SU-DW}
	\\
\raisebox{-.2\height}{
 	\begin{tikzpicture}
 	\tikzset{node distance = 1.5cm}
	\tikzstyle{gauge} = [circle, draw,inner sep=3pt];
	\tikzstyle{flavour} = [regular polygon,regular polygon sides=4,inner 
sep=3pt, draw];
%
\node (g1) [flavour, label= left:{$\scriptscriptstyle{SU(N{+}1)}$}]{};
\node (g2) [flavour,below of=g1,label= 
left:{$\scriptscriptstyle{USp(2N{-}2)}$}] {};
%
%
 	\tikzset{node distance = 0.75cm}
    \node (Lup) [above of= g1] {};
    \node (Ldown) [below of= g2] {};
\draw (g1)--(g2) (g1)--(Lup) (g2)--(Ldown);
	\end{tikzpicture}
	}
\begin{tikzpicture}[rotate=-90,scale=0.5]
\fill[top color=gray!50!black,bottom color=gray!10,middle 
color=gray,shading=axis,opacity=0.10] (0,0) circle (2cm and 0.5cm);
\fill[left color=gray!50!black,right color=gray!50!black,middle 
color=gray!50,shading=axis,opacity=0.10] (2,0) -- (2,6) arc (360:180:2cm and 
0.5cm) -- (-2,0) arc (180:360:2cm and 0.5cm);
\fill[top color=gray!90!,bottom color=gray!2,middle 
color=gray!30,shading=axis,opacity=0.10] (0,6) circle (2cm and 0.5cm);
\draw (-2,6) -- (-2,0) arc (180:360:2cm and 0.5cm) -- (2,6) ++ (-2,0) circle 
(2cm and 0.5cm);
\draw[densely dashed] (-2,0) arc (180:0:2cm and 0.5cm);
\end{tikzpicture}
\raisebox{-.2\height}{
 	\begin{tikzpicture}
 	\tikzset{node distance = 1.5cm}
	\tikzstyle{gauge} = [circle, draw,inner sep=3pt];
	\tikzstyle{flavour} = [regular polygon,regular polygon sides=4,inner 
sep=3pt, draw];
%
\node (g1) [flavour, label= right:{$\scriptscriptstyle{SO(2N{+}2)}$}]{};
\node (g2) [flavour,below of=g1,label= 
right:{$\scriptscriptstyle{SU(N{+}1)}$}] {};
%
%
 	\tikzset{node distance = 0.75cm}
    \node (Lup) [above of= g1] {};
    \node (Ldown) [below of= g2] {};
\draw (g1)--(g2) (g1)--(Lup) (g2)--(Ldown);
	\end{tikzpicture}
	}
		\label{eq:mixed-DW}
\end{align}
\end{subequations}
In order to obtain the desired $4$d theories, the vector 
multiplets have to be added back, i.e.\ they gauge the global symmetries 
associated to the punctures. Therefore, the $4$d theories on a torus can be 
constructed via gluing the fundamental domain walls along their punctures. For 
example, a torus theory constructed from four domain walls can be sketched as 
in Figure \ref{fig:T4DW}.
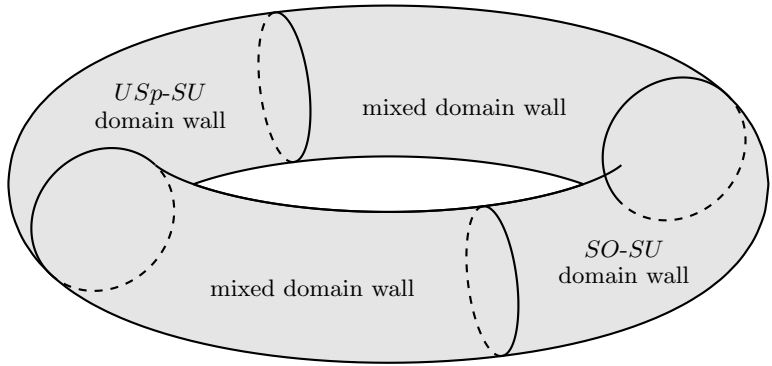
\begin{figure}[t]
\centering
\tdplotsetmaincoords{70}{0}
\tikzset{declare function={torusx(\u,\v,\R,\r)=cos(\u)*(\R + \r*cos(\v)); 
torusy(\u,\v,\R,\r)=(\R + \r*cos(\v))*sin(\u);
torusz(\u,\v,\R,\r)=\r*sin(\v);
vcrit1(\u,\th)=atan(tan(\th)*sin(\u));
vcrit2(\u,\th)=180+atan(tan(\th)*sin(\u));
disc(\th,\R,\r)=((pow(\r,2)-pow(\R,2))*pow(cot(\th),2)+%
pow(\r,2)*(2+pow(tan(\th),2)))/pow(\R,2);
umax(\th,\R,\r)=ifthenelse(disc(\th,\R,\r)>0,asin(sqrt(abs(disc(\th,\R,\r)))),
0);
}}
\begin{tikzpicture}[tdplot_main_coords]
\pgfmathsetmacro{\R}{4}
\pgfmathsetmacro{\r}{1}
 \draw[thick,fill=gray,even odd rule,fill opacity=0.2] 
plot[variable=\x,domain=0:360,smooth,samples=71]
 ({torusx(\x,vcrit1(\x,\tdplotmaintheta),\R,\r)},
 {torusy(\x,vcrit1(\x,\tdplotmaintheta),\R,\r)},
 {torusz(\x,vcrit1(\x,\tdplotmaintheta),\R,\r)}) 
 plot[variable=\x, 
domain={-180+umax(\tdplotmaintheta,\R,\r)}:{-umax(\tdplotmaintheta,\R,\r)},
smooth,samples=51]
 ({torusx(\x,vcrit2(\x,\tdplotmaintheta),\R,\r)},
 {torusy(\x,vcrit2(\x,\tdplotmaintheta),\R,\r)},
 {torusz(\x,vcrit2(\x,\tdplotmaintheta),\R,\r)})
 plot[variable=\x,
domain={umax(\tdplotmaintheta,\R,\r)}:{180-umax(\tdplotmaintheta,\R,\r)},smooth,
samples=51]
 ({torusx(\x,vcrit2(\x,\tdplotmaintheta),\R,\r)},
 {torusy(\x,vcrit2(\x,\tdplotmaintheta),\R,\r)},
 {torusz(\x,vcrit2(\x,\tdplotmaintheta),\R,\r)});
 \draw[thick] plot[variable=\x,
domain={-180+umax(\tdplotmaintheta,\R,\r)/2}:{-umax(\tdplotmaintheta,\R,\r)/2},
smooth,samples=51]
 ({torusx(\x,vcrit2(\x,\tdplotmaintheta),\R,\r)},
 {torusy(\x,vcrit2(\x,\tdplotmaintheta),\R,\r)},
 {torusz(\x,vcrit2(\x,\tdplotmaintheta),\R,\r)});
 \foreach \X  in {20,110,200,290}  
 {\draw[thick,dashed]
plot[smooth,variable=\x,domain={360+vcrit1(\X,\tdplotmaintheta)}:{vcrit2(\X,
\tdplotmaintheta)},samples=71]   
 ({torusx(\X,\x,\R,\r)},{torusy(\X,\x,\R,\r)},{torusz(\X,\x,\R,\r)});
 \draw[thick]
plot[smooth,variable=\x,domain={vcrit2(\X,\tdplotmaintheta)}:{vcrit1(\X,
\tdplotmaintheta)},samples=71]   
 ({torusx(\X,\x,\R,\r)},{torusy(\X,\x,\R,\r)},{torusz(\X,\x,\R,\r)})
 ;
 }
\draw (1,3) node {\footnotesize{mixed domain wall}};
\draw (-1,-4) node {\footnotesize{mixed domain wall}};
\draw (3.1,-2.5) node {\footnotesize{$SO$-$SU$}};
\draw (3.1,-3.5) node {\footnotesize{domain wall}};
\draw (-3,3.5) node {\footnotesize{$USp$-$SU$}};
\draw (-3,2.5) node {\footnotesize{domain wall}};
\end{tikzpicture}
\caption{Example of a $4$d theory on a torus constructed from a 
 $USp$-$SU$ domain wall \eqref{eq:Sp-SU-DW} and a $SO$-$SU$ domain wall 
\eqref{eq:SO-SU-DW} such that both are glued to two mixed domain walls 
\eqref{eq:mixed-DW}.}
\label{fig:T4DW}
\end{figure}
Nevertheless, all these fundamental domain walls potentially suffer from 
various anomalies. In particular, one has to guarantee that the glued 
$SU(N\pm 1)$ gauge nodes are free of the cubic gauge anomalies, i.e.\
\begin{align}
\Tr\left(SU(N\pm 1)^3\right)=0 \,.
\end{align}
These gauge anomalies are the focus of the next section.
\subsection{\texorpdfstring{$\Tr\,(SU(N\pm 1)^3)$}{Tr(SU3)} cubic gauge 
anomalies}
Since all three fundamental domain walls \eqref{eq:DW_sketch} contain unitary 
gauge nodes, they will 
be discussed separately. Appendix \ref{app:anomalies} provides the conventions 
for the anomaly coefficients. Without loss of generality, one may consider the 
$k=1$ case, i.e.\ the  $5$d theory 
\eqref{eq:5d_quiver_affine_A} consists only of one $USp(2N{-}2)$ and one 
$SO(2N{+}2)$ node, to construct the domain walls. This $k=1$ is only used to 
simplify the presentation, but any comparison to $6$d requires $k\geq2$. 
\paragraph{$\boldsymbol{USp(2N{-}2)}$-$\boldsymbol{SU(N{+}1)}$ domain walls.}
To begin with, focus on the domain wall \eqref{eq:Sp-SU-DW} and $k=1$. 
Consequently, each chamber of the domain wall contains one 
$USp(2N{-}2)$ and one $SU(N{+}1)$ node. 
To connect the chambers, one includes 
additional $4$d chiral fields $q$ to each node, which are oriented from left 
to right. Then there are two choices for which nodes between the two chambers 
the chirals $q$ connect: either the $USp$-$SU$ and $SU$-$USp$ nodes, or 
$USp$-$USp$ and $SU$-$SU$ nodes. In addition, for each choice, there exist four 
types of boundary conditions that can be assigned on the $5$d hypermultiplets 
of the two chambers. Following Section \ref{sec:5d_half_BPS_bc}, these boundary 
conditions are denoted by $(+,+)$, $(+,-)$, $(-,+)$, and $(-,-)$. 
If one chooses to connect the $USp$-$SU$ and $SU$-$USp$ nodes in the two 
chambers, then the boundary conditions for the $5$d fields of left and right 
chamber have to be opposite to each other.
As a consequence, the boundary conditions of the $5$d fields together with 
the choice of chiral fields $q$ determine how to include additional $4$d 
chiral 
fields $\tilde q$ that serve to formulate cubic superpotentials and 
triangulate the domain walls.
In total there are eight different 
$USp(2N{-}2)$-$SU(N{+}1)$ domain walls for the $k=1$ case. 
To illustrate the construction, the fundamental domain wall with $(+,-)$ 
boundary conditions on the left and $(-,+)$ on the right is given by
\begin{align}
	\raisebox{-.5\height}{
 	\begin{tikzpicture}
 	\tikzset{node distance = 2cm}
	\tikzstyle{gauge} = [circle, draw,inner sep=3pt];
	\tikzstyle{flavour} = [regular polygon,regular polygon sides=4,inner 
sep=3pt, draw];
%
\node (g1) [flavour,label= left:{$\scriptscriptstyle{SU(N{+}1)}$}] 
{};
\node (g2) [flavour,right of=g1, 
label= right:{$\scriptscriptstyle{USp(2N{-}2)}$}] {};
\node (g3) [flavour, above of 
=g2,label= right:{$\scriptscriptstyle{SU(N{+}1)}$}] 
{};
\node (g4) [flavour, left 
of=g3,label= left:{$\scriptscriptstyle{USp(2N{-}2)}$}] 
{};
%
%
 	\tikzset{node distance = 1cm}
    \node (Lup) [above of= g4] {};
    \node (Ldown) [below of= g1] {};
    \node (Rup) [above of= g3] {};
    \node (Rdown) [below of= g2] {};
    \node (Cup) [right of=Lup] {};
    \node (Cdown) [right of=Ldown] {};
    \begin{scope}[decoration={markings,mark =at position 0.5 with 
{\arrow{stealth}}}]
    \draw[postaction={decorate}] (Ldown) -- (g1);
    \draw[postaction={decorate}] (g1) -- (g4);
    \draw[postaction={decorate}] (g4) -- (Lup);
    \draw[postaction={decorate}] (Rdown) -- (g2);
    \draw[postaction={decorate}] (g2) -- (g3);
    \draw[postaction={decorate}] (g3) -- (Rup);
    \draw[postaction={decorate}] (g1) -- (g2);
    \draw[postaction={decorate}] (g4) -- (g3);
    \draw[postaction={decorate}] (g3) -- (g1);
    \draw[postaction={decorate}] (Cup) -- (g4);
    \draw[postaction={decorate}] (g2) -- (Cdown);
     \end{scope}
%
\draw (g1)  to[in=-205,out=-245,loop] (g1);
\draw (-0.55,0.4) node {$\scriptscriptstyle{A}$};
\draw (g3)  to[in=-295,out=-335,loop] (g3);
\draw (2+0.55,2+0.4) node {$\scriptscriptstyle{\bar{A}}$};
%
\draw (1,-0.2) node {$\scriptscriptstyle{q_{2}}$};
\draw (1,1-0.25) node {$\scriptscriptstyle{\tilde{q}_{1}}$};
\draw (1,2-0.2) node {$\scriptscriptstyle{q_{1}}$};
\draw (1,3-0.25) node {$\scriptscriptstyle{\tilde{q}_{2}}$};
	\end{tikzpicture}
	}
	\label{eq:USpSU+-}
\end{align}
and the need for the additional loops $A$ and $\bar{A}$ 
can be seen as follows:
From $4$d perspective, \eqref{eq:USpSU+-} is an \none Wess-Zumino model with 
non-Abelian global symmetries $USp(2N{-}2)^2\times SU(N{+}1)^2$.
In order to glue this domain wall with others, one further needs to require 
that the cubic gauge anomaly of $SU(N{+}1)$ nodes vanishes. For 
example, consider the lower left $SU(N{+}1)$ node to be specific. 
Then the $4$d chiral fields $q_2$ and $\tilde q_1$ contribute 
$-(2N-2)$ and $(N+1)$ units to the cubic anomaly, respectively. The 
two vertical chiral fields, as residues of the $5$d hypermultiplets, also contribute to cubic anomaly with an additional factor of $\frac{1}{2}$, as a consequence of $5$d anomaly inflow, see \cite{Kim:2018lfo}. In this quiver, the net cubic anomaly from the two vertical chirals is zero due to our chosen boundary condition. Summing up all the contributions, the cubic anomaly becomes
\begin{align}
\Tr\left(SU(N+ 1)^3\right)
=-(2N-2)+(N+1)
=-(N+1-4)\,.
\end{align}
Therefore, one additional anti-symmetric chiral matter fields $A$ is required to 
cancel the anomaly of the $SU(N{+}1)$ node. 

Analogously, one can show that the upper right $SU(N{+}1)$ node in 
\eqref{eq:USpSU+-} requires another matter fields $\bar A$ in the conjugate representation of the anti-symmetric 
representation to cancel the cubic gauge anomaly. 
Thus, the domain wall becomes anomaly free if the additional loops $A$ and $\bar{A}$ are added to the construction as shown in 
\eqref{eq:USpSU+-}.

From the quiver diagram, it is straightforward to derive the \none 
superpotential. Besides the cubic superpotential terms that stem from
triangles in \eqref{eq:USpSU+-}, one needs also form superpotentials for the  
anti-symmetric fields $A$ and $\bar{A}$ via
\begin{align}
\mc W=\Tr\,(J q_1 \bar A q_1+J q_2 A q_2)
+\text{cubic terms}\,,
\label{P2}
\end{align}
where $J$ is the anti-symmetric tensor of $USp(2N{-}2)$.
%
%
\paragraph{$\boldsymbol{SO(2N{+}2)}$-$\boldsymbol{SU(N{-}1)}$ domain walls.}
Analogous to the above analysis, there exist eight different 
$SO$-$SU$ domains walls \eqref{eq:SO-SU-DW} in the $k=1$ case, which follow from 
the four types of boundary conditions assigned to the $5$d fields and the 
subsequent two choices of how to connect the $5$d chambers via additional $4$d 
chiral fields.
As an example, consider the domain wall with boundary conditions $(+,-)_{L,R}$ 
such that the $SO$-$SU$ and $SU$-$SO$ nodes are paired up. In order to have 
vanishing cubic gauge anomalies for the $SU(N{-}1)$ nodes, one has to add one 
symmetric matter field for each $SU(N{-}1)$ node. Collecting all the 
ingredients, one ends up with the following theory:
\begin{align}
	\raisebox{-.5\height}{
 	\begin{tikzpicture}
 	\tikzset{node distance = 2cm}
	\tikzstyle{gauge} = [circle, draw,inner sep=3pt];
	\tikzstyle{flavour} = [regular polygon,regular polygon sides=4,inner 
sep=3pt, draw];
%
\node (g1) [flavour,label= left:{$\scriptscriptstyle{SU(N{-}1)}$}] 
{};
\node (g2) [flavour,right of=g1, label= right:{$\scriptscriptstyle{SO(2N{+}2)}$}] 
{};
\node (g3) [flavour, above of =g2,label= 
right:{$\scriptscriptstyle{SU(N{-}1)}$}] 
{};
\node (g4) [flavour, left of=g3,label= left:{$\scriptscriptstyle{SO(2N{+}2)}$}] 
{};
%
%
 	\tikzset{node distance = 1cm}
    \node (Lup) [above of= g4] {};
    \node (Ldown) [below of= g1] {};
    \node (Rup) [above of= g3] {};
    \node (Rdown) [below of= g2] {};
    \node (Cup) [right of=Lup] {};
    \node (Cdown) [right of=Ldown] {};
    \begin{scope}[decoration={markings,mark =at position 0.5 with 
{\arrow{stealth}}}]
    \draw[postaction={decorate}] (Ldown) -- (g1);
    \draw[postaction={decorate}] (g1) -- (g4);
    \draw[postaction={decorate}] (g4) -- (Lup);
    \draw[postaction={decorate}] (Rdown) -- (g2);
    \draw[postaction={decorate}] (g2) -- (g3);
    \draw[postaction={decorate}] (g3) -- (Rup);
    \draw[postaction={decorate}] (g1) -- (g2);
    \draw[postaction={decorate}] (g4) -- (g3);
    \draw[postaction={decorate}] (g3) -- (g1);
    \draw[postaction={decorate}] (Cup) -- (g4);
    \draw[postaction={decorate}] (g2) -- (Cdown);
     \end{scope}
%
\draw (g1)  to[in=-205,out=-245,loop] (g1);
\draw (-0.50,0.35) node {$\scriptscriptstyle{S}$};
\draw (g3)  to[in=-295,out=-335,loop] (g3);
\draw (2+0.55,2+0.4) node {$\scriptscriptstyle{\bar{S}}$};
	\end{tikzpicture}
	}
	\label{eq:SOSU+-}
\end{align}
It is then straightforward to verify the vanishing cubic 
anomalies for the two $SU(N{-}1)$ nodes. Moreover, the superpotential 
can be derived in similar fashion as in the $USp$-$SU$ case.
%
%
\paragraph{$\boldsymbol{USp(2N{-}2)}$-$\boldsymbol{SO(2N{+}2)}$-$\boldsymbol{
SU(N{\pm}1)}$ mixed domain walls.}
Lastly, consider the mixed domain wall \eqref{eq:mixed-DW} as the interface 
theory of a 
$USp(2N{-}2)$-$SU(N{+}1)$ and a $SO(2N{+}2)$-$SU(N{-}1)$ chamber. 
Unfortunately, it turns out that there is no way to 
construct a cubic anomaly free mixed domain wall for certain choices of 
boundary conditions. 
For instance, consider the mixed domain wall corresponding to 
$(+,-)_L$, $(-,+)_R$ boundary conditions, which is given by
\begin{align}
	\raisebox{-.5\height}{
 	\begin{tikzpicture}
 	\tikzset{node distance = 2cm}
	\tikzstyle{gauge} = [circle, draw,inner sep=3pt];
	\tikzstyle{flavour} = [regular polygon,regular polygon sides=4,inner 
sep=3pt, draw];
%
\node (g1) [flavour,label= left:{$\scriptscriptstyle{SU(N{-}1)}$}] 
{};
\node (g2) [flavour,right of=g1, label= 
right:{$\scriptscriptstyle{USp(2N{-}2)}$}] {};
\node (g3) [flavour, above of =g2,label= 
right:{$\scriptscriptstyle{SU(N{+}1)}$}] 
{};
\node (g4) [flavour, left of=g3,label= left:{$\scriptscriptstyle{SO(2N{+}2)}$}] 
{};
%
%
 	\tikzset{node distance = 1cm}
    \node (Lup) [above of= g4] {};
    \node (Ldown) [below of= g1] {};
    \node (Rup) [above of= g3] {};
    \node (Rdown) [below of= g2] {};
    \node (Cup) [right of=Lup] {};
    \node (Cdown) [right of=Ldown] {};
    \begin{scope}[decoration={markings,mark =at position 0.5 with 
{\arrow{stealth}}}]
    \draw[postaction={decorate}] (Ldown) -- (g1);
    \draw[postaction={decorate}] (g1) -- (g4);
    \draw[postaction={decorate}] (g4) -- (Lup);
    \draw[postaction={decorate}] (Rdown) -- (g2);
    \draw[postaction={decorate}] (g2) -- (g3);
    \draw[postaction={decorate}] (g3) -- (Rup);
    \draw[postaction={decorate}] (g1) -- (g2);
    \draw[postaction={decorate}] (g4) -- (g3);
    \draw[postaction={decorate}] (g3) -- (g1);
    \draw[postaction={decorate}] (Cup) -- (g4);
    \draw[postaction={decorate}] (g2) -- (Cdown);
     \end{scope}
%
\draw (g1)  to[in=-25,out=-65,loop] (g1);
\draw (0.5,-0.35) node {$\scriptscriptstyle{A}$};
\draw (g3)  to[in=-205,out=-245,loop] (g3);
\draw (2-0.5,2+0.35) node {$\scriptscriptstyle{\bar{S}}$};
%
	\end{tikzpicture}
	}
	\label{eq:USpSOSU+-}
\end{align}
One can check that the $SU(N\pm 1)$ nodes, even when supplemented with  
(anti-)symmetric matter fields, are anomalous, i.e.\
\begin{subequations}
\begin{align}
\Tr\,(SU(N{+}1)^3)&=(2N+2)-(N-1)-(N+1+4)=-2\,, \\
\Tr\,(SU(N{-}1)^3)&=-(2N-2)+(N+1)+(N-1-4)=-2\,.
\end{align} 
\end{subequations}
Of course, the anomalous domain wall cannot be glued to other anomaly 
free domain walls. 
Nevertheless, it is possible to find other domain walls with anomalies of 
opposite and glue these consistently. For example, one may glue the 
anomalous domain wall \eqref{eq:USpSOSU+-} with 
another anomalous domain wall of the form
\begin{align}
	\raisebox{-.5\height}{
 	\begin{tikzpicture}
 	\tikzset{node distance = 2cm}
	\tikzstyle{gauge} = [circle, draw,inner sep=3pt];
	\tikzstyle{flavour} = [regular polygon,regular polygon sides=4,inner 
sep=3pt, draw];
%
\node (g1) [flavour,label= left:{$\scriptscriptstyle{USp(2N{-}2)}$}] 
{};
\node (g2) [flavour,right of=g1, label= 
right:{$\scriptscriptstyle{SU(N{-}1)}$}] {};
\node (g3) [flavour, above of =g2,label= 
right:{$\scriptscriptstyle{SO(2N{+}2)}$}] 
{};
\node (g4) [flavour, left of=g3,label= 
left:{$\scriptscriptstyle{SU(N{+}1)}$}] 
{};
%
%
 	\tikzset{node distance = 1cm}
    \node (Lup) [above of= g4] {};
    \node (Ldown) [below of= g1] {};
    \node (Rup) [above of= g3] {};
    \node (Rdown) [below of= g2] {};
    \node (Cup) [right of=Lup] {};
    \node (Cdown) [right of=Ldown] {};
    \begin{scope}[decoration={markings,mark =at position 0.5 with 
{\arrow{stealth}}}]
    \draw[postaction={decorate}] (Ldown) -- (g1);
    \draw[postaction={decorate}] (g1) -- (g4);
    \draw[postaction={decorate}] (g4) -- (Lup);
    \draw[postaction={decorate}] (Rdown) -- (g2);
    \draw[postaction={decorate}] (g2) -- (g3);
    \draw[postaction={decorate}] (g3) -- (Rup);
    \draw[postaction={decorate}] (g1) -- (g2);
    \draw[postaction={decorate}] (g4) -- (g3);
    \draw[postaction={decorate}] (g3) -- (g1);
    \draw[postaction={decorate}] (Cup) -- (g4);
    \draw[postaction={decorate}] (g2) -- (Cdown);
     \end{scope}
%
\draw (g2)  to[in=-205,out=-245,loop] (g2);
\draw (2-0.5,+0.35) node {$\scriptscriptstyle{\bar{A}}$};
\draw (g4)  to[in=-25,out=-65,loop] (g4);
\draw (0.5,2-0.35) node {$\scriptscriptstyle{S}$};
%
	\end{tikzpicture}
	}
	\label{eq:USpSOSU-+}
\end{align}
which has $+2$ cubic $SU(N\pm 1)$ anomalies.
A calculation proves that the two mixed domain 
walls \eqref{eq:USpSOSU+-} and \eqref{eq:USpSOSU-+} can be glued 
to form an anomaly free $4$d theory on a torus given by the quiver
\begin{align}
	\raisebox{-.5\height}{
 	\begin{tikzpicture}
 	\tikzset{node distance = 2cm}
	\tikzstyle{gauge} = [circle, draw,inner sep=3pt];
	\tikzstyle{flavour} = [regular polygon,regular polygon sides=4,inner 
sep=3pt, draw];
%
\node (g1) [gauge,label= above left:{$\scriptscriptstyle{SU(N{-}1)}$}] 
{};
\node (g2) [gauge,right of=g1, 
label= below right:{$\scriptscriptstyle{USp(2N{-}2)}$}] {};
\node (g3) [gauge, above of 
=g2,label= below right:{$\scriptscriptstyle{SU(N{+}1)}$}] 
{};
\node (g4) [gauge, left 
of=g3,label= above left:{$\scriptscriptstyle{SO(2N{+}2)}$}] 
{};
%
%
 	\tikzset{node distance = 1cm}
    \node (Lup) [above of= g4] {};
    \node (Ldown) [below of= g1] {};
    \node (Left1) [left of= g4] {};
    \node (Left2) [below of= Left1,label=right:{$\scriptscriptstyle{c}$}] {};
    \node (Left3) [below of= Left2] {};
    \node (Rup) [above of= g3] {};
    \node (Right1) [right of= g3] {};
    \node (Right2) [below of= Right1,label=left:{$\scriptscriptstyle{c}$}] {};
    \node (Right3) [below of= Right2] {};
    \node (Rdown) [below of= g2] {};
    \node (Cup) [right of=Lup,label=below:{$\scriptscriptstyle{a}$}] {};
    \node (Cdown) [right of=Ldown,label=above:{$\scriptscriptstyle{a}$}] {};
    \node (Dup) [right of=Rup,label=below:{$\scriptscriptstyle{b}$}] {};
    \node (Ddown) [left of=Ldown,label=above:{$\scriptscriptstyle{b}$}] {};
    \begin{scope}[decoration={markings,mark =at position 0.5 with 
{\arrow{stealth}}}]
    \draw[postaction={decorate}] (Ldown) -- (g1);
    \draw[postaction={decorate}] (g1) -- (g4);
    \draw[postaction={decorate}] (g4) -- (Lup);
    \draw[postaction={decorate}] (Rdown) -- (g2);
    \draw[postaction={decorate}] (g2) -- (g3);
    \draw[postaction={decorate}] (g3) -- (Rup);
    \draw[postaction={decorate}] (g1) -- (g2);
    \draw[postaction={decorate}] (g4) -- (g3);
    \draw[postaction={decorate}] (g3) -- (g1);
    \draw[postaction={decorate}] (Cup) -- (g4);
    \draw[postaction={decorate}] (g2) -- (Cdown);
    \draw[postaction={decorate}] (Left1) -- (g4);
    \draw[postaction={decorate}] (Left3) -- (g1);
    \draw[postaction={decorate}] (g4)--(Left2);
    \draw[postaction={decorate}] (g3) -- (Right1);
    \draw[postaction={decorate}] (g2) -- (Right3);
    \draw[postaction={decorate}] (Right2)--(g2);
    \draw[postaction={decorate}] (g1) -- (Ddown);
    \draw[postaction={decorate}] (Dup)--(g3);
     \end{scope}
%
\draw (g1)  to[in=-50,out=-80,loop] (g1);
\draw (0.35,-0.50) node {$\scriptscriptstyle{A}$};
\draw (g1)  to[in=-10,out=-40,loop] (g1);
\draw (0.60,-0.20) node {$\scriptscriptstyle{\bar{A}}$};
\draw (g3)  to[in=-190,out=-220,loop] (g3);
\draw (2-0.35,2+0.50) node {$\scriptscriptstyle{S}$};
\draw (g3)  to[in=-230,out=-260,loop] (g3);
\draw (2-0.6,2+0.2) node {$\scriptscriptstyle{\bar{S}}$};
%
	\end{tikzpicture}
	}
	\label{eq:USpSOSUonT}
\end{align}
However, since all the gauge nodes have positive $\beta$-functions one would 
expect that the theory \eqref{eq:USpSOSUonT} is IR free.
Moreover, the $U(1)_R$-charge anomalies of \eqref{eq:USpSOSUonT} fail to match 
with the $6$d predictions.
Therefore, the next section focuses on more constraints in order to screen out 
possible $4$d candidate theories on a torus.
\subsection{\texorpdfstring{$\Tr\,(U(1)_RG^2)$}{Tr(RGG)} anomalies}
Next consider the $U(1)_R$-current anomalies. The $U(1)_R$ symmetry is 
the maximal torus of the $SU(2)_R$ inherited from the $6$d/$5$d origin and, 
thus, not the genuine $R$-symmetry of the corresponding $4$d SCFT in the IR. 
In this paper, the assumption is that this $U(1)_R$ symmetry is preserved in  
the domain walls as well as the $4$d quiver theories on torus. One expects that 
the $U(1)_R$ mixes with other global $U(1)$ flavour symmetries. Therefore, the 
genuine IR $R$-symmetry can be determined via $a$-maximisation.

Since the UV $U(1)_R$-symmetry is the maximal torus of the $SU(2)_R$,
the hypermultiplets in the $5$d quiver theory \eqref{eq:5d_quiver_affine_A} 
can be normalised to have $U(1)_R$-charge $1$. 
As a consequence, the inherited (vertical) chiral fields of the domain wall 
theories have unity $U(1)_R$-charge too. Following \cite{Gaiotto:2015una, Kim:2018lfo, Kim:2018bpg}, one may 
assign $U(1)_R$-charge $0$ to the horizontal chiral matter fields $q$. Then,  
the $U(1)_R$-charges of all remaining chiral superfields are determined via the
superpotential. The $U(1)_R$-charge assignment for the quiver theories used in 
this paper are summarised in Table \ref{tab:Rcharge}.
\begin{table}[t]
\centering
\begin{tabular}{c|ccc}
\toprule 
Fields & vertical/diagonal & horizontal & (anti-)symmetric \\ \midrule
$U(1)_R$ & 1 & 0 & 2 \\ 	
\bottomrule
\end{tabular}
\caption{$U(1)_R$-charge assignments for chiral superfields in the domain 
walls.}
\label{tab:Rcharge}
\end{table}
With the $U(1)_R$-charge at hand, one can analyse the potential 
$\Tr\,(U(1)_R G^2)$ anomalies that may prevent a consistent gluing of 
various domains walls via 
gauging the corresponding non-Abelian $G=USp,SO,SU$ groups. 
Note that one needs to add a $\mathcal N=1$ vector multiplet in order to gauge 
a non-Abelian gauge group $G$, and the associated gauginos have $U(1)_R$-charge 
$1$, due to the $5$d origin. 
The contributions to $\Tr\,(U(1)_RG^2)$ from the vertical chiral superfields 
and the gauginos should be treated as $5$d anomaly inflow, as before, and, 
thus, appear with an additional $\frac{1}{2}$ prefactor. From $4$d 
perspective, this $\frac{1}{2}$ factor can be interpreted as avoiding double 
counting of the common gaugino and chiral multiplet contribution when gluing two 
domain walls.   

Due to the assumption that the glued domain walls and 
further quiver theories on torus 
preserve the $U(1)_R$ symmetry, it is natural to impose the 
constraint
\begin{align}
\Tr\,\left( U(1)_RG^2 \right)=0\,,
\label{eq:trRG2}
\end{align}
for all $G=USp,SO,SU$ groups. However, a computations show that none of the 
previous cubic-anomaly-free domain wall satisfies the constraint 
\eqref{eq:trRG2}.
For instance, consider the domain wall \eqref{eq:USpSU+-}, then one computes 
the anomaly
\begin{align}
\Tr\,\left( U(1)_R\,USp(2N{-}2)^2 
\right)=(0-1)\cdot(N+1)+\frac{1}{2}(2N-2+2)=-1 
\label{eq:trRUSP2}
\end{align}
for both $USp(2N{-}2)$ nodes in the quiver. The non-zero anomaly signals 
that this domain wall cannot be glued with another $U(1)_R$-anomaly-free domain 
wall. 
Nonetheless, one may try to weaken the constraint \eqref{eq:trRG2}: If there 
would exist a domain wall with $\Tr\,(U(1)_R\,USp(2N{-}2)^2)=+1$ anomaly, one 
could glue these two anomalous domain walls while preserving the 
$U(1)_R$ symmetry in the resulting theory. 

Based on the above analysis, one can exhaust all possible domain walls for 
a given $k$ with various boundary conditions, compute their cubic gauge and 
$\Tr\,(U(1)_RG^2)$ anomalies, and then combine those with both opposite cubic 
anomalies for $SU$ nodes, and opposite $\Tr\,(U(1)_RG^2)$ anomalies for all 
$G=USp,SO,SU$ nodes. 
However, it turns out that no domain wall pairing can 
simultaneously satisfy both the cubic and the $\Tr\,(U(1)_RG^2)$ anomaly 
conditions. 
In fact, the vertical and diagonal chiral fields with $R$-charge $1$ do not 
contribute to $\Tr\,(U(1)_RG^2)$. Thus, the $\Tr\,(U(1)_RG^2)$ values for each 
$G=USp,SO,SU$ node of a given type of domain walls is 
always independent on the boundary conditions imposed. 
The implication seems to be that one cannot preserve the $U(1)_R\subset SU(2)_R$ 
symmetry with $R$-charge assignments of Table \ref{tab:Rcharge} for the domain 
walls constructed from $USp$, $SO$, $SU$ punctures. 

One possibility to resolve the matter is to assign different $R$-charges to 
all chiral fields. Then, the cubic gauge  and $\Tr\,(U(1)_RG^2)$ anomalies 
serve as constraints to determine possible $R$-charge assignments. 
There are two scenarios to consider: On the one hand, assume that 
the $U(1)_R\subset SU(2)_R$ is preserved during the  domain 
wall construction. Then the vertical chiral fields, which are induced from 
$5$d hypermultiplets, will inherit the $U(1)_R$-charge as 
before, i.e.\ $U(1)_R$-charge equal to $1$. Assigning different  
$U(1)_R$-charges to the horizontal chiral multiplets determines the $R$-charges 
of all other chiral fields. However, a computations shows that 
there does not exist an $R$-charge assignment that satisfies all three 
$\Tr\,(U(1)_RG^2)$ anomaly constraints simultaneously. 

On the other hand, one may assume 
that the $U(1)_R\subset SU(2)_R$ is broken in the domain wall construction. 
Then the broken $U(1)_R$ symmetry could be mixed with another 
broken $U(1)$ flavour symmetry to form a new $U(1)^\prime_R$ symmetry.
This would allow to assign $U(1)^\prime_R$-charges different from $1$ to the 
vertical chiral fields.
Subsequently, the new freedom allows for a charge assignment compatible with 
the vanishing of all $\Tr\,(U(1)_R^\prime G^2)$ anomalies. 

Nevertheless, 
choosing to work with the $U(1)^\prime_R$ symmetry means that one can no longer 
compare to the $4$d results obtained from compactifications of the $6$d anomaly 
polynomial.
For example, the $4$d anomaly polynomial 
\eqref{eq:4d_anomaly_torus} imposes two additional constraints from vanishing 
$U(1)_R$-gravity 
anomalies, $\Tr\,(U(1)_R)=\Tr\,(U(1)_R^3)=0$. Moreover, there are predictions 
from $6$d for various $\Tr\,(U(1)_R 
U(1)_{F_i} U(1)_{F_j})$ anomalies, 
all of which have to be matched with any $4$d candidate theory.
Therefore, in this paper it is assumed that the $U(1)_R\subset SU(2)_R$ 
symmetry is unbroken in the domain walls and quiver theories on torus. 
To circumvent the problem of non-vanishing $\Tr\,SU^3$ and $\Tr\,(U(1)_R G^2)$ 
anomalies, 
the 
domain wall theories are changed by considering \emph{non-maximal} boundary 
conditions on the vector and hypermultiplets of the $5$d quiver theory. The 
domain wall theories obtained via the non-maximal boundary conditions have 
puncture symmetries that are smaller rank subgroups of 
$USp(2N{-}2)$ and $SO(2N{+}2)$, see also  
\cite{Gaiotto:2015una}.
%
%
\subsection{Non-maximal boundary conditions}
\label{sec:non-max_bc}
The necessity to introduce domain walls with non-maximal boundary 
conditions can be traced back to \eqref{eq:trRUSP2}. 
If there would be a domain wall with 
$SU(N)$-$USp(2N{-}2)$ punctures, then the $\Tr\,(U(1)_R G^2)$ vanishes for 
both $G=SU(N)$ and $USp(2N{-}2)$. 
To achieve that one has to assign non-maximal 
boundary conditions for the hypermultiplets as well as gauge multiplets correspondingly. 
\paragraph{Boundary conditions for domain wall.}
The general idea to assign non-maximal boundary conditions is as follows: 
For a $USp(2N{-}2)$ gauge node with $(N+1)$ attached 
hypermultiplets, for instance, one notices from the $5$d quiver 
\eqref{eq:5d_quiver_affine_A} that these $(N+1)$ hypermuliplets can be treated 
as $(2N{+}2)$ half-hypermultiplets, or say chirals, and their global symmetry is 
gauged by a $5$d $\mc N=1$ $SO(2N{+}2)$ gauge multiplet. 
If one removes this $SO(2N{+}2)$ gauge multiplet, the half-hypers would at most 
have a $SU(2N+2)$ flavor symmetry. 
Now, when two $5$d theories are placed at the interface, for example at 
$x_4=0$, to construct a domain wall, one can assign mixed boundary conditions to 
the $(2N+2)$ half-hypermultiplets. Generalising the discussion of Section 
\ref{sec:5d_half_BPS_bc}, some number of chiral multiplets are assigned 
Dirichlet boundary conditions, while the remaining fields obey Neumann boundary 
conditions such that only a smaller flavour group $H\subset SU(2N+2)$ is 
preserved.  
As long as $H$ can be embedded into $SO(2N+2)$, one 
can add back part of the $SO(2N{+}2)$ gauge multiplet to gauge this flavour 
symmetry $H$. 
The mixed boundary conditions imposed on the hypermultiplets always 
preserve half of the $5$d $\mc N=1$ supersymmetries, just like the
$\tfrac{1}{2}$ BPS conditions do. 
On the other hand, one assigns Neumann boundary 
conditions to the gauge fields componetns in the subgroup $H$, and 
Dirichlet boundary conditions to the rest of gauge degrees of 
freedom. The adjoint chiral multiplet in the $5$d $\mc N=1$ $SO(2N+2)$ gauge 
multiplet is chosen to obey Dirichlet boundary conditions. In total, the 
conditions are
\begin{align}
\label{eq:non-max_bc_domain-wall}
\begin{cases}
 A^{SO(2N{+}2)}_4\vert_{x^4=0}= 0= 
 \partial_4 A^{SO(2N{+}2)}_\mu\vert_{x^4=0} \\
A^{SO(2N{+}2)\backslash H}_{\mu} \vert_{x^4=0}=0 \\
\Phi^{SO(2N{+}2)}\vert_{x^4=0}= 0
\end{cases}
\for \mu=0,1,2,3\,.
\end{align}
Obviously, such boundary condition on the gauge multiplet turn out to be 
$\tfrac{1}{2}$ BPS and as such preserve half of $5$d $\mc N=1$ 
supersymmetries. 
\paragraph{Boundary conditions for punctures.}
So far, non-maximal boundary conditions have been discussed for the domain 
wall. However, one needs to defined the boundary conditions for punctures with 
non-maximal symmetry as well. 
Suppose the construction contains a non-maximal open puncture with symmetry 
group $H$, then, as customary in the literature 
\cite{Kim:2018lfo,Razamat:2018gro}, one reverses the above boundary conditions 
on the gauge multiplet, i.e.\
\begin{align}
\label{eq:non-max_bc_puncture}
\begin{cases}
 A^{SO(2N{+}2)}_\mu\vert_{x^4=0}= 0 \\
\Phi^{SO(2N+2)\backslash H}\vert_{x^4=0}= 0=\partial_4\Phi^{H\subset SO(2N{+}2)}\vert_{x^4=0}
\end{cases}
\for \mu=0,1,2,3,4\,.
\end{align}
Such non-maximal boundary conditions imply that the punctures have a reduced 
global symmetry $H$. The non-zero components of 
the adjoint chiral $\Phi^{H}$ induces a negative anomaly 
inflow contribution from the $5$d quiver theory to the $4$d domain wall due to 
the $H$-puncture. Similar non-maximal boundary conditions can be also applied 
to the $SO(2N{+}2)$ gauge nodes in the $5$d quiver 
\eqref{eq:5d_quiver_affine_A} with $(N-1)$ attached hypermultiplets. This 
becomes relevant at the end of this section.

\paragraph{$\boldsymbol{SU(N)}$-$\boldsymbol{USp(2N{-}2)}$ domain wall.}
In order to construct a domain wall with $SU(N)$-$USp(2N{-}2)$ puncture 
symmetry, in contrast to the maximal boundary conditions \eqref{eq:USp2} and 
\eqref{eq:USp2max}, one only 
imposes $\tfrac{1}{2}$ BPS boundary conditions for the first $N$ 
hypermultiplets of the $USp(2N{-}2)$ node
\begin{subequations}
\label{eq:USpsmall1}
\begin{align}
 +) \quad \partial_4 X_i\vert_{x^4=0} &=Y_i\vert_{x^4=0}=0\,,\ 
\quad \text{or} \quad    
-)\quad
X_i\vert_{x^4=0}=\partial_4 Y_i\vert_{x^4=0}=0\,,\quad \text{for} \quad 
i=1,2,\dots,N \,,
\end{align}
while the remaining hypermultiplets are forced to 
vanish, i.e.
\begin{align}
 X_{N+1} &=Y_{N+1}=0\,.
\end{align} 
\end{subequations}
Correspondingly, one also has to reduce the rank of gauge group 
via suitable vector multiplet boundary conditions,
\begin{align}
\begin{cases}
 A^{SO(2N{+}2)}_4\vert_{x^4=0}= 0= 
 \partial_4 A^{SO(2N{+}2)}_\mu\vert_{x^4=0} \\
A^{SO(2N{+}2)\backslash SU(N)}_{\mu} \vert_{x^4=0}=0
\end{cases}
\for \mu=0,1,2,3\,.
\label{eq:USpsmall2} 
\end{align}
The adjoint chiral field is subjected to Dirichlet boundary conditions as 
above.

Starting, for instance, from \eqref{eq:USpSU+-}, these non-maximal 
boundary conditions allow to obtain the following 
modified domain wall with $SU(N)$-$USp(2N{-}2)$ punctures:
\begin{align}
	\raisebox{-.5\height}{
 	\begin{tikzpicture}
 	\tikzset{node distance = 2cm}
	\tikzstyle{gauge} = [circle, draw,inner sep=3pt];
	\tikzstyle{flavour} = [regular polygon,regular polygon sides=4,inner 
sep=3pt, draw];
%
\node (g1) [flavour,label= left:{$\scriptscriptstyle{SU(N)}$}] 
{};
\node (g2) [flavour,right of=g1, 
label= right:{$\scriptscriptstyle{USp(2N{-}2)}$}] {};
\node (g3) [flavour, above of 
=g2,label= right:{$\scriptscriptstyle{SU(N)}$}] 
{};
\node (g4) [flavour, left 
of=g3,label= left:{$\scriptscriptstyle{USp(2N{-}2)}$}] 
{};
%
%
 	\tikzset{node distance = 1cm}
    \node (Lup) [above of= g4] {};
    \node (Ldown) [below of= g1] {};
    \node (Rup) [above of= g3] {};
    \node (Rdown) [below of= g2] {};
    \node (Cup) [right of=Lup] {};
    \node (Cdown) [right of=Ldown] {};
    \begin{scope}[decoration={markings,mark =at position 0.5 with 
{\arrow{stealth}}}]
    \draw[postaction={decorate}] (Ldown) -- (g1);
    \draw[postaction={decorate}] (g1) -- (g4);
    \draw[postaction={decorate}] (g4) -- (Lup);
    \draw[postaction={decorate}] (Rdown) -- (g2);
    \draw[postaction={decorate}] (g2) -- (g3);
    \draw[postaction={decorate}] (g3) -- (Rup);
    \draw[postaction={decorate}] (g1) -- (g2);
    \draw[postaction={decorate}] (g4) -- (g3);
    \draw[postaction={decorate}] (g3) -- (g1);
    \draw[postaction={decorate}] (Cup) -- (g4);
    \draw[postaction={decorate}] (g2) -- (Cdown);
     \end{scope}
%
\draw (g1)  to[in=-205,out=-245,loop] (g1);
\draw (-0.55,0.4) node {$\scriptscriptstyle{A}$};
\draw (g3)  to[in=-295,out=-335,loop] (g3);
\draw (2+0.55,2+0.4) node {$\scriptscriptstyle{\bar{A}}$};
%
\draw (1,-0.2) node {$\scriptscriptstyle{q_{2}}$};
\draw (1,1-0.25) node {$\scriptscriptstyle{\tilde{q}_{1}}$};
\draw (1,2-0.2) node {$\scriptscriptstyle{q_{1}}$};
\draw (1,3-0.25) node {$\scriptscriptstyle{\tilde{q}_{2}}$};
	\end{tikzpicture}
	}
	\label{eq:USpSU}
\end{align}
One can verify explicitly that all $\Tr\,(U(1)_R G^2)$ anomalies vanish. 
Of course, with this modification, one would worry about the cubic 
anomaly of the $SU(N)$ nodes. However, as explained in previous sections, one 
can always glue two cubic anomalous domain walls with opposite cubic anomalies 
to render the $SU(N)$ node anomaly free.
\paragraph{$\boldsymbol{SO(2N{+}2)}$-$\boldsymbol{SU(N{-}1)}$ domain wall.}
Similarly for a $SO(2N{+}2)$-$SU(N{-}1)$ domain wall, see for example 
\eqref{eq:SOSU+-}, one computes a non-vanishing 
$\Tr\,(U(1)_R G^2)$ anomaly for $G=SO(2N{+}2)$, i.e.
\begin{align}
\Tr\,\left(U(1)_R SO(2N{+}2)^2 \right)=(0-1)\cdot(N-1)+\frac{1}{2}(2N+2-2)=+1\,.
\end{align}
If one considers a $SO(2N{+}2)$-$SU(N)$ domain wall instead, one finds
\begin{align}
\Tr\,\left(U(1)_R SO(2N{+}2)^2 \right)=0= \Tr\,\left(U(1)_R SU(N)^2 \right)\,.
\end{align}
However, $SU(N)$ cannot be embedded into $USp(2N{-}2)$. This suggests to 
consider a \emph{smaller puncture} by assigning non-maximal $\frac{1}{2}$ BPS 
boundary conditions on the vector multiplet for only a subgroup 
$SO(2N)\subset SO(2N{+}2)$,
\begin{align}
\begin{cases}
 A^{SO(2N{+}2)}_4\vert_{x^4=0}= 0 =\partial_4 
A^{SO(2N{+}2)}_\mu\vert_{x^4=0} \\
A^{SO(2N{+}2)\backslash SO(2N)}_{\mu} \vert_{x^4=0}=0
\end{cases}
\for \mu=0,1,2,3\,,
\label{eq:SOsmall} 
\end{align}
such that a domain wall with puncture symmetry $SO(2N)$-$SU(N{-}1)$ arises, i.e.
\begin{align}
	\raisebox{-.5\height}{
 	\begin{tikzpicture}
 	\tikzset{node distance = 2cm}
	\tikzstyle{gauge} = [circle, draw,inner sep=3pt];
	\tikzstyle{flavour} = [regular polygon,regular polygon sides=4,inner 
sep=3pt, draw];
%
\node (g1) [flavour,label= left:{$\scriptscriptstyle{SU(N{-}1)}$}] 
{};
\node (g2) [flavour,right of=g1, label= right:{$\scriptscriptstyle{SO(2N)}$}] 
{};
\node (g3) [flavour, above of =g2,label= 
right:{$\scriptscriptstyle{SU(N{-1})}$}] 
{};
\node (g4) [flavour, left of=g3,label= left:{$\scriptscriptstyle{SO(2N)}$}] 
{};
%
%
 	\tikzset{node distance = 1cm}
    \node (Lup) [above of= g4] {};
    \node (Ldown) [below of= g1] {};
    \node (Rup) [above of= g3] {};
    \node (Rdown) [below of= g2] {};
    \node (Cup) [right of=Lup] {};
    \node (Cdown) [right of=Ldown] {};
    \begin{scope}[decoration={markings,mark =at position 0.5 with 
{\arrow{stealth}}}]
    \draw[postaction={decorate}] (Ldown) -- (g1);
    \draw[postaction={decorate}] (g1) -- (g4);
    \draw[postaction={decorate}] (g4) -- (Lup);
    \draw[postaction={decorate}] (Rdown) -- (g2);
    \draw[postaction={decorate}] (g2) -- (g3);
    \draw[postaction={decorate}] (g3) -- (Rup);
    \draw[postaction={decorate}] (g1) -- (g2);
    \draw[postaction={decorate}] (g4) -- (g3);
    \draw[postaction={decorate}] (g3) -- (g1);
    \draw[postaction={decorate}] (Cup) -- (g4);
    \draw[postaction={decorate}] (g2) -- (Cdown);
     \end{scope}
%
\draw (g1)  to[in=-205,out=-245,loop] (g1);
\draw (-0.50,0.35) node {$\scriptscriptstyle{S}$};
\draw (g3)  to[in=-295,out=-335,loop] (g3);
\draw (2+0.55,2+0.4) node {$\scriptscriptstyle{\bar{S}}$};
	\end{tikzpicture}
	}
	\label{eq:SOSU}
\end{align}
For completeness, \eqref{eq:SOsmall} is supplemented by Dirichlet boundary 
condition for the adjoint-valued chiral in the $SO(2N+2)$ vector multiplet. 
\paragraph{Mixed domain wall.}
Lastly, for the mixed domain wall \eqref{eq:USpSOSU-+}, one finds 
that the vanishing $\Tr\,(U(1)_R G^2)$ anomaly constraints for all $G=USp,SO,SU$ 
require to further reduce 
the $USp(2N{-}2)-SU(N)$ puncture symmetry of the quiver \eqref{eq:USpSOSU+-}
to $USp(2N{-}4)-SU(N{-}1)$. To do 
so, on the left chamber, one assigns
$\frac{1}{2}$ BPS boundary conditions on the vector multiplet for only a 
subgroup $USp(2N{-}4)\subset USp(2N{-}2)$, i.e.\
\begin{subequations}
\label{eq:USpsmallmix}
\begin{align}
 &\begin{cases}
  A^{USp(2N{-}2)}_4\vert_{x^4=0}=0 = \partial_4 
A^{USp(2N{-}2)}_\mu\vert_{x^4=0} 
\\
A^{USp(2N{-}2)\backslash 
USp(2N{-}4)}_{\mu} \vert_{x^4=0}=0 \\
\Phi^{USp(2N{-}2)}\vert_{x^4=0}= 0
\end{cases}
\,, \for \mu=0,1,2,3 \,,
\\
\notag\\
&\begin{cases}
  A^{SO(2N{+}2)}_4 \vert_{x^4=0} =0
=\partial_4 A^{SO(2N{+}2)}_{\mu} \vert_{x^4=0} \\
A^{SO(2N{+}2)\backslash SU(N{-}1)}_{\mu}  \vert_{x^4=0}=0 \\
\Phi^{SO(2N{+}2)}\vert_{x^4=0}= 0
 \end{cases}
\,,\for \mu=0,1,2,3\,.
\end{align}
In addition, $\frac{1}{2}$ BPS 
boundary conditions are imposed for the first $(N-1)$ 
hypermultiplets
\begin{align}
 +) \quad \partial_4 X_i\vert_{x^4=0}=Y_i\vert_{x^4=0}=0\,,\ 
\quad \text{or} \quad    
-)\quad
X_i\vert_{x^4=0}=\partial_4 Y_i\vert_{x^4=0}=0\,,\quad \text{for }\, 
i=1,2,\dots,N-1 \,,
\end{align}
while Dirichlet boundary conditions are imposed on the 
remaining two hypermultiplets
\begin{align}
X_{N}=Y_{N}=X_{N+1}=Y_{N+1}=0\,.
\end{align} 
\end{subequations}
On the right chamber, the boundary conditions are assigned to be the same as in
\eqref{eq:SOsmall}. The resulting domain wall is
\begin{align}
	\raisebox{-.5\height}{
 	\begin{tikzpicture}
 	\tikzset{node distance = 2cm}
	\tikzstyle{gauge} = [circle, draw,inner sep=3pt];
	\tikzstyle{flavour} = [regular polygon,regular polygon sides=4,inner 
sep=3pt, draw];
%
\node (g1) [flavour,label= left:{$\scriptscriptstyle{USp(2N{-}4)}$}] 
{};
\node (g2) [flavour,right of=g1, label= 
right:{$\scriptscriptstyle{SU(N{-}1)}$}] {};
\node (g3) [flavour, above of =g2,label= 
right:{$\scriptscriptstyle{SO(2N)}$}] 
{};
\node (g4) [flavour, left of=g3,label= 
left:{$\scriptscriptstyle{SU(N{-}1)}$}] 
{};
%
%
 	\tikzset{node distance = 1cm}
    \node (Lup) [above of= g4] {};
    \node (Ldown) [below of= g1] {};
    \node (Rup) [above of= g3] {};
    \node (Rdown) [below of= g2] {};
    \node (Cup) [right of=Lup] {};
    \node (Cdown) [right of=Ldown] {};
    \begin{scope}[decoration={markings,mark =at position 0.5 with 
{\arrow{stealth}}}]
    \draw[postaction={decorate}] (Ldown) -- (g1);
    \draw[postaction={decorate}] (g1) -- (g4);
    \draw[postaction={decorate}] (g4) -- (Lup);
    \draw[postaction={decorate}] (Rdown) -- (g2);
    \draw[postaction={decorate}] (g2) -- (g3);
    \draw[postaction={decorate}] (g3) -- (Rup);
    \draw[postaction={decorate}] (g1) -- (g2);
    \draw[postaction={decorate}] (g4) -- (g3);
    \draw[postaction={decorate}] (g3) -- (g1);
    \draw[postaction={decorate}] (Cup) -- (g4);
    \draw[postaction={decorate}] (g2) -- (Cdown);
     \end{scope}
%
\draw (g2)  to[in=-205,out=-245,loop] (g2);
\draw (2-0.5,+0.35) node {$\scriptscriptstyle{\bar{A}}$};
\draw (g4)  to[in=-25,out=-65,loop] (g4);
\draw (0.5,2-0.35) node {$\scriptscriptstyle{S}$};
%
	\end{tikzpicture}
	}
	\label{eq:USpSOSU}
\end{align}
In total, there are still three types of domain walls, but the  puncture 
symmetries have changed to $USp(2N{-}4)$-$SU(N{-}1)$ and $SO(2N)$-$SU(N{-}1)$. 
These 
domain walls are equipped with a non-anomalous $U(1)_R\subset SU(2)_R$ 
symmetry, i.e.
\begin{align}
\Tr\,\left(U(1)_R USp(2N{-}4)^2 \right)=
\Tr\,\left(U(1)_R SO(2N)^2 \right)=
\Tr\,\left(U(1)_R SU(N{-}1)^2 \right)=0\,,
\end{align}
with $U(1)_R$-charge assignments as in Table \ref{tab:Rcharge}. 
In addition, computing the $U(1)_R$-gravity anomalies reveals that
\begin{align}
\Tr\,\left(U(1)_R \right)=0=\Tr\,\left( U(1)_R^3 \right)\,,
\end{align}
holds for all three domain walls. 
It implies that for a quiver theory, constructed from glueing these 
domain walls on a torus, the $U(1)_R$ symmetry is preserved and the
$U(1)_R$-gravity anomalies vanish. 
Consequently, there is a possibility that such $4$d quiver 
theories have a $6$d origin. 
Despite the intermediate success of constructing sensible theories via domain 
walls, the necessity of non-maximal boundary conditions opens the door to a 
partial loss of information from the $6$d origin.
In the next section, the $4$d theories constructed from 
non-maximal boundary conditions will be considered in detail.
%
%
\section{Four dimensions}
\label{sec:4d}
In this section, the $4$d theories are constructed via domain walls that arise 
from non-maximal boundary conditions.
\subsection{Domain wall for non-maximal boundary conditions}
Here an example made of $USp$-mixed-$SO$-mixed domain walls is presented, which 
satisfies the criteria 
\begin{align}
 \Tr \left( U(1)_R \right)= 0 =  \Tr \left( U(1)_R^3 \right)
\end{align}
with non-zero $a$-central charge.
\paragraph{$\boldsymbol{USp(2N{-}4)}$-$\boldsymbol{SU(N{-}1)}$ domain wall.} 
Collecting all the ingredients, the domain wall becomes
\begin{align}
	\raisebox{-.5\height}{
 	\begin{tikzpicture}
 	\tikzset{node distance = 2cm}
	\tikzstyle{gauge} = [circle, draw,inner sep=3pt];
	\tikzstyle{flavour} = [regular polygon,regular polygon sides=4,inner 
sep=3pt, draw];
%
\node (g1) [flavour,label=left:{$\scriptscriptstyle{SU(N{-}1)}$}] 
{};
\node (g2) [flavour,right of=g1, 
label=right:{$\scriptscriptstyle{USp(2N{-}4)}$}] {};
\node (g3) [flavour, above of 
=g2,label=right:{$\scriptscriptstyle{SU(N{-}1)}$}] 
{};
\node (g4) [flavour, left 
of=g3,label=left:{$\scriptscriptstyle{USp(2N{-}4)}$}] 
{};
%
%
 	\tikzset{node distance = 1cm}
    \node (Lup) [above of= g4] {};
    \node (Ldown) [below of= g1] {};
    \node (Rup) [above of= g3] {};
    \node (Rdown) [below of= g2] {};
    \node (Cup) [right of=Lup,label=below:{$\scriptscriptstyle{a}$}] {};
    \node (Cdown) [right of=Ldown,label=above:{$\scriptscriptstyle{a}$}] {};
    \begin{scope}[decoration={markings,mark =at position 0.5 with 
{\arrow{stealth}}}]
    \draw[postaction={decorate}] (Ldown) -- (g1);
    \draw[postaction={decorate}] (g1) -- (g4);
    \draw[postaction={decorate}] (g4) -- (Lup);
    \draw[postaction={decorate}] (Rdown) -- (g2);
    \draw[postaction={decorate}] (g2) -- (g3);
    \draw[postaction={decorate}] (g3) -- (Rup);
    \draw[postaction={decorate}] (g1) -- (g2);
    \draw[postaction={decorate}] (g4) -- (g3);
    \draw[postaction={decorate}] (g3) -- (g1);
    \draw[postaction={decorate}] (Cup) -- (g4);
    \draw[postaction={decorate}] (g2) -- (Cdown);
     \end{scope}
%
\draw (g1)  to[in=-120,out=-160,loop] (g1);
\draw (-0.55,-0.35) node {$\scriptscriptstyle{A}$};
\draw (g3)  to[in=-20,out=-60,loop] (g3);
\draw (2.55,2-0.35) node {$\scriptscriptstyle{\bar{A}}$};
%
	\end{tikzpicture}
	}
	\label{eq:SpDW}
\end{align}
where $A$, $\bar{A}$ denote chiral supermultiplets in the anti-symmetric 
representation and its conjugate.
Being careful, one should analyse various anomalies. Firstly, 
for the cubic anomalies of the left and right $SU(N)$ groups, one finds
\begin{subequations}
\begin{align}
\Tr(SU(N)_L^3) &=-(2N-4)+(N-1)+(N-5)=-2 \,,\\
\Tr(SU(N)_R^3) &=+(2N-4)-(N-1)+(N-5)=+2 \,.
\end{align}
\end{subequations}
Secondly, to compute $U(1)_R$ anomalies of the four nodes, one needs to clarify 
the $R$-charge assignment. Following the discussion of the last sections, i.e.\ 
the $R$-charge of  all vertical lines is $1$, and horizontal lines have 
$R$-charge 
$0$. Hence, the superpotential imposes that diagonal lines and 
(anti-)symmetric matter fields have their $R$-charges 
$1$ and $2$, respectively. 
Likewise, the gauge nodes contribute as $\frac{1}{2}$ of the $4$d gauginos' 
anomaly contribution with $R$-charge $1$, due to $5$d anomaly inflow. Then, 
one computes
\begin{subequations}
\begin{align}
\Tr(U(1)_R\,USp(2N{-}4)_L^2) &=(0-1)\cdot (N-1)+\frac{1}{2}\cdot 2(N-1)=0 \,,\\
\Tr(U(1)_R\,SU(N{-}1)_L^2) &=(0-1)\cdot (2N-4)+(2-1)\cdot(N-3)+\frac{1}{2}\cdot 
2(N-1)=0 \,,\\
\Tr(U(1)_R\,USp(2N{-}4)_R^2) &=(0-1)\cdot (N-1)+\frac{1}{2}\cdot 2(N-1)=0 \,, \\
\Tr(U(1)_R\,SU(N{-}1)_R^2) &=(0-1)\cdot (2N-4)+(2-1)\cdot(N-4)+\frac{1}{2}\cdot 
2(N-1)=0\,.
\end{align}
\end{subequations}
Lastly, computing the $U(1)_R$-gravity anomalies yields
\begin{align}
\Tr\left(U(1)_R \right)=\Tr\left(U(1)_R^3\right)=2\cdot \bigg( &-(2N-4)\cdot 
(N-1)+\frac{1}{2}(N-1)(N-2) \notag \\
&+\frac{1}{2}((N-1)^2-1)+\frac{1}{2}(N-2)(2N-3)\bigg)=0\,,
\end{align}
where the gaugino contribution is again scaled by a factor of a half.

After these considerations, one can now proceed to glue 
the domain wall \eqref{eq:SpDW} from the right or left to some other domain 
walls with cubic anomalies $+2$/$-2$ on the left or right side. 
In order to do so, one needs to introduce other types of domain walls.
\paragraph{Mixed domain wall.}
The mixed domain wall, meaning that all three types $SU$, $USp$, $SO$ of groups 
are present, is defined as follows:
\begin{align}
	\raisebox{-.5\height}{
 	\begin{tikzpicture}
 	\tikzset{node distance = 2cm}
	\tikzstyle{gauge} = [circle, draw,inner sep=3pt];
	\tikzstyle{flavour} = [regular polygon,regular polygon sides=4,inner 
sep=3pt, draw];
%
\node (g1) [flavour,label=left:{$\scriptscriptstyle{USp(2N{-}4)}$}] 
{};
\node (g2) [flavour,right of=g1, 
label=right:{$\scriptscriptstyle{SU(N{-}1)}$}] {};
\node (g3) [flavour, above of 
=g2,label=right:{$\scriptscriptstyle{SO(2N)}$}] 
{};
\node (g4) [flavour, left 
of=g3,label=left:{$\scriptscriptstyle{SU(N{-}1)}$}] 
{};
%
%
 	\tikzset{node distance = 1cm}
    \node (Lup) [above of= g4] {};
    \node (Ldown) [below of= g1] {};
    \node (Rup) [above of= g3] {};
    \node (Rdown) [below of= g2] {};
    \node (Cup) [right of=Lup,label=below:{$\scriptscriptstyle{b}$}] {};
    \node (Cdown) [right of=Ldown,label=above:{$\scriptscriptstyle{b}$}] {};
    \begin{scope}[decoration={markings,mark =at position 0.5 with 
{\arrow{stealth}}}]
    \draw[postaction={decorate}] (g1)--(Ldown);
    \draw[postaction={decorate}] (g4) -- (g1);
    \draw[postaction={decorate}] (Lup)--(g4);
    \draw[postaction={decorate}] (g2)--(Rdown);
    \draw[postaction={decorate}] (g3) -- (g2);
    \draw[postaction={decorate}] (Rup)--(g3);
    \draw[postaction={decorate}] (g2) -- (g1);
    \draw[postaction={decorate}] (g3) -- (g4);
    \draw[postaction={decorate}] (g1) -- (g3);
    \draw[postaction={decorate}] (g4)--(Cup);
    \draw[postaction={decorate}] (Cdown)--(g2);
     \end{scope}
%
\draw (g4)  to[in=-120,out=-160,loop] (g4);
\draw (-0.55,-0.35+2) node {$\scriptscriptstyle{\bar{S}}$};
\draw (g2)  to[in=-20,out=-60,loop] (g2);
\draw (2.55,2-0.35-2) node {$\scriptscriptstyle{A}$};
%
	\end{tikzpicture}
	}
	\label{eq:mixDW}
\end{align}
where $\bar{S}$ is the conjugate of the symmetric representation.
Recall that the additional loops have been introduced so that these two domain 
walls can be $S$-glued \cite{Razamat:2016dpl}. $S$-gluing makes all chiral 
fields corresponding to vertical lines massive and, thus, they can be integrated 
out. In particular, this will guarantee  that all $SO(2N)$ and $USp(2N{-}4)$ 
nodes 
are asymptotically free. The computations of various anomalies is analogous to 
the above cases, such that one readily obtains
\begin{subequations}
\begin{align}
\Tr(SU(N{-}1)_L^3) &=+(2N)-(N-1)-(N+3)=-2 \,, \\
\Tr(SU(N{-}1)_R^3) &=-(2N-4)+(N-1)+(N-5)=-2 \,, \\
\Tr(U(1)_R\,USp(2N{-}4)_L^2) &=(0-1)\cdot (N-1)+\frac{1}{2}\cdot 2(N-1)=0 \,,\\
\Tr(U(1)_R\,SU(N{-}1)_L^2) &=(0-1)\cdot 2N+(2-1)\cdot(N+1)+\frac{1}{2}\cdot 
2(N-1)=0  \,,\\
\Tr(U(1)_R\,SO(2N)_R^2) &=(0-1)\cdot (N-1)+\frac{1}{2}\cdot 2(N-1)=0 \,, \\
\Tr(U(1)_R\,SU(N{-}1)_R^2) &=(0-1)\cdot (2N-4)+(2-1)\cdot(N-3)+\frac{1}{2}\cdot 
2(N-1) \notag \\
& =0 \,,\\
\Tr\left(U(1)_R \right)=\Tr\left(U(1)_R^3 \right) 
&=-(2N-4)(N-1)-2N(N-1)+\frac{1}{2}(N-1)(N-2) \notag \\
&\quad +\frac{1}{2}N(N-1) +2\cdot \frac{1}{2}((N-1)^2-1)
+\frac{1}{2}(N-2)(2N-3) \notag \\
&\quad +\frac{1}{2}N(2N-1)  \notag \\
&=0\,.
\end{align}
\end{subequations}
Clearly the domain wall \eqref{eq:mixDW} with cubic anomaly $-2$ can be glued 
to the first $USp$-$SU$ domain wall \eqref{eq:SpDW} from the right side. Note 
in particular that it also a conformal gluing with respect to the 
$U(1)_R$ symmetry.
Next, one has to construct additional domain walls, for instance a $SO$-$SU$ 
and another mixed domain wall, such that the domain walls \eqref{eq:SpDW}, 
\eqref{eq:mixDW} can be glued to close the torus.
\paragraph{$SO$-$SU$ and mixed domain walls.}
The remaining possibilities are as follows:
\begin{align}
	\raisebox{-.5\height}{
 	\begin{tikzpicture}
 	\tikzset{node distance = 2cm}
	\tikzstyle{gauge} = [circle, draw,inner sep=3pt];
	\tikzstyle{flavour} = [regular polygon,regular polygon sides=4,inner 
sep=3pt, draw];
%
\node (g1) [flavour,label=left:{$\scriptscriptstyle{SU(N{-}1)}$}] 
{};
\node (g2) [flavour,right of=g1, 
label=right:{$\scriptscriptstyle{SO(2N)}$}] {};
\node (g3) [flavour, above of 
=g2,label=right:{$\scriptscriptstyle{SU(N{-}1)}$}] {};
\node (g4) [flavour, left 
of=g3,label=left:{$\scriptscriptstyle{SO(2N)}$}] 
{};
%
%
 	\tikzset{node distance = 1cm}
    \node (Lup) [above of= g4] {};
    \node (Ldown) [below of= g1] {};
    \node (Rup) [above of= g3] {};
    \node (Rdown) [below of= g2] {};
    \node (Cup) [right of=Lup,label=below:{$\scriptscriptstyle{c}$}] {};
    \node (Cdown) [right of=Ldown,label=above:{$\scriptscriptstyle{c}$}] {};
    \begin{scope}[decoration={markings,mark =at position 0.5 with 
{\arrow{stealth}}}]
    \draw[postaction={decorate}] (Ldown) -- (g1);
    \draw[postaction={decorate}] (g1) -- (g4);
    \draw[postaction={decorate}] (g4) -- (Lup);
    \draw[postaction={decorate}] (Rdown) -- (g2);
    \draw[postaction={decorate}] (g2) -- (g3);
    \draw[postaction={decorate}] (g3) -- (Rup);
    \draw[postaction={decorate}] (g1) -- (g2);
    \draw[postaction={decorate}] (g4) -- (g3);
    \draw[postaction={decorate}] (g3) -- (g1);
    \draw[postaction={decorate}] (Cup) -- (g4);
    \draw[postaction={decorate}] (g2) -- (Cdown);
     \end{scope}
%
\draw (g1)  to[in=-120,out=-160,loop] (g1);
\draw (-0.55,-0.35) node {$\scriptscriptstyle{S}$};
\draw (g3)  to[in=-20,out=-60,loop] (g3);
\draw (2.55,2-0.35) node {$\scriptscriptstyle{\bar{S}}$};
%
	\end{tikzpicture}
	}
\qquad 
	\raisebox{-.5\height}{
 	\begin{tikzpicture}
 	\tikzset{node distance = 2cm}
	\tikzstyle{gauge} = [circle, draw,inner sep=3pt];
	\tikzstyle{flavour} = [regular polygon,regular polygon sides=4,inner 
sep=3pt, draw];
%
\node (g1) [flavour,label=left:{$\scriptscriptstyle{SO(2N)}$}] 
{};
\node (g2) [flavour,right of=g1, 
label=right:{$\scriptscriptstyle{SU(N{-}1)}$}] {};
\node (g3) [flavour, above of 
=g2,label=right:{$\scriptscriptstyle{USp(2N{-}4)}$}] 
{};
\node (g4) [flavour, left 
of=g3,label=left:{$\scriptscriptstyle{SU(N{-}1)}$}] 
{};
%
%
 	\tikzset{node distance = 1cm}
    \node (Lup) [above of= g4] {};
    \node (Ldown) [below of= g1] {};
    \node (Rup) [above of= g3] {};
    \node (Rdown) [below of= g2] {};
    \node (Cup) [right of=Lup,label=below:{$\scriptscriptstyle{d}$}] {};
    \node (Cdown) [right of=Ldown,label=above:{$\scriptscriptstyle{d}$}] {};
    \begin{scope}[decoration={markings,mark =at position 0.5 with 
{\arrow{stealth}}}]
    \draw[postaction={decorate}] (g1)--(Ldown);
    \draw[postaction={decorate}] (g4) -- (g1);
    \draw[postaction={decorate}] (Lup)--(g4);
    \draw[postaction={decorate}] (g2)--(Rdown);
    \draw[postaction={decorate}] (g3) -- (g2);
    \draw[postaction={decorate}] (Rup)--(g3);
    \draw[postaction={decorate}] (g2) -- (g1);
    \draw[postaction={decorate}] (g3) -- (g4);
    \draw[postaction={decorate}] (g1) -- (g3);
    \draw[postaction={decorate}] (g4)--(Cup);
    \draw[postaction={decorate}] (Cdown)--(g2);
     \end{scope}
%
\draw (g4)  to[in=-120,out=-160,loop] (g4);
\draw (-0.55,-0.35+2) node {$\scriptscriptstyle{\bar{A}}$};
\draw (g2)  to[in=-20,out=-60,loop] (g2);
\draw (2.55,2-0.35-2) node {$\scriptscriptstyle{S}$};
%
	\end{tikzpicture}
	}
	\label{eq:rest_DW}
\end{align}
For these two domain walls, one can easily compute the anomalies as before.  
The $SO$-$SU$ domain wall has $+2$ and $-2$ cubic anomaly for the left and 
right $SU$ gauge node, respectively. The mixed domain wall one has $+2$ cubic 
anomaly for both $SU$ gauge nodes. All other anomalies 
vanishes for both domain walls. 
Therefore, the four domain walls \eqref{eq:SpDW}, \eqref{eq:mixDW}, 
\eqref{eq:rest_DW} can be conformally glued successively to form a quiver 
theory on a torus.
\subsection{Quiver theory on torus}
\label{sec:quiver_torus}
\begin{figure}[t]
\centering
\begin{tikzpicture}
 	\tikzset{node distance = 2cm}
	\tikzstyle{gauge} = [circle, draw,inner sep=3pt];
	\tikzstyle{flavour} = [regular polygon,regular polygon sides=4,inner 
sep=3pt, draw];
%
\node (g8) [gauge,label=below:{$\scriptscriptstyle{SU(N{-}1)}$}] 
{$\scriptscriptstyle{8}$};
\node (g5) [gauge,right of=g8, 
label=below :{$\scriptscriptstyle{USp(2N{-}4)}$}] 
{$\scriptscriptstyle{5}$};
\node (g6) [gauge, right of=g5,label=below:{$\scriptscriptstyle{SU(N{-}1)}$}] 
{$\scriptscriptstyle{6}$};
\node (g7) [gauge, right of=g6,label=below:{$\scriptscriptstyle{SO(2N)}$}] 
{$\scriptscriptstyle{7}$};
\node (g4) [gauge, above right of 
=g8,label=above:{$\scriptscriptstyle{SU(N{-}1)}$}] 
{$\scriptscriptstyle{4}$};
\node (g1) [gauge, right of=g4,label=above:{$\scriptscriptstyle{SO(2N)}$}] 
{$\scriptscriptstyle{1}$};
\node (g2) [gauge,right of=g1,label=above:{$\scriptscriptstyle{SU(N{-}1)}$}] 
{$\scriptscriptstyle{2}$};
\node (g3) [gauge, right of=g2,label=above:{$\scriptscriptstyle{USp(2N{-}4)}$}] 
{$\scriptscriptstyle{3}$};
%
%
\node(ddown) [below left of= g8,label=above:{$\scriptscriptstyle{d}$}] {};
\node(adown) [below left of= g5,label=above:{$\scriptscriptstyle{a}$}] {};
\node(bdown) [below left of= g6,label=above:{$\scriptscriptstyle{b}$}] {};
\node(cdown) [below left of= g7,label=above:{$\scriptscriptstyle{c}$}] {};
\node(bup) [above right of= g4,label=left:{$\scriptscriptstyle{b}$}] {};
\node(cup) [above right of= g1,label=left:{$\scriptscriptstyle{c}$}] {};
\node(dup) [above right of= g2,label=left:{$\scriptscriptstyle{d}$}] {};
\node(aup) [above right of= g3,label=left:{$\scriptscriptstyle{a}$}] {};
    \node(Lup) [left of= g4] {};
    \node(Ldown) [left of= g8] {};
    \node(Rup) [right of= g3] {};
    \node(Rdown) [right of= g7] {};
    \begin{scope}[decoration={markings,mark =at position 0.5 with 
{\arrow{stealth}}}]
    \draw[postaction={decorate}] (g8) -- (g5);
    \draw[postaction={decorate}] (g8) -- (g5);
    \draw[postaction={decorate}] (g6) -- (g5);
    \draw[postaction={decorate}] (g6) -- (g7);
    \draw[postaction={decorate}] (g1) -- (g4);
    \draw[postaction={decorate}] (g1) -- (g2);
    \draw[postaction={decorate}] (g3) -- (g2);
    \draw[postaction={decorate}] (g4) -- (g8);
    \draw[postaction={decorate}] (g5) -- (g1);
    \draw[postaction={decorate}] (g2) -- (g6);
    \draw[postaction={decorate}] (g7) -- (g3);
    \draw[postaction={decorate}] (ddown)--(g8);
    \draw[postaction={decorate}] (g5)--(adown);
    \draw[postaction={decorate}] (bdown)--(g6);
    \draw[postaction={decorate}] (g7)--(cdown);
    \draw[postaction={decorate}] (g4)--(bup);
    \draw[postaction={decorate}] (cup)--(g1);
    \draw[postaction={decorate}] (g2)--(dup);
    \draw[postaction={decorate}] (aup)--(g3);
    \draw[postaction={decorate}] (Lup)--(g4);
    \draw[postaction={decorate}] (g8)--(Ldown);
    \draw[postaction={decorate}] (g3)--(Rup);
    \draw[postaction={decorate}] (Rdown)--(g7);
     \end{scope}
%
\draw (g4)  to[in=-80,out=-120,loop] (g4);
\draw (g4)  to[in=-20,out=-60,loop] (g4);
\draw (2,1) node {$\scriptscriptstyle{\bar{S}}$};
\draw (1.5,0.8) node {$\scriptscriptstyle{\bar{A}}$};
\draw (g2)  to[in=-80,out=-120,loop] (g2);
\draw (g2)  to[in=-20,out=-60,loop] (g2);
\draw (2+4,1) node {$\scriptscriptstyle{\bar{S}}$};
\draw (1.5+4,0.8) node {$\scriptscriptstyle{\bar{A}}$};
\draw (g8)  to[in=80+45,out=120+45,loop] (g8);
\draw (g8)  to[in=20+45,out=60+45,loop] (g8);
\draw (0,0.75) node {$\scriptscriptstyle{S}$};
\draw (-0.5,0.55) node {$\scriptscriptstyle{A}$};
\draw (g6)  to[in=80+45,out=120+45,loop] (g6);
\draw (g6)  to[in=20+45,out=60+45,loop] (g6);
\draw (0+4,0.75) node {$\scriptscriptstyle{S}$};
\draw (-0.5+4,0.55) node {$\scriptscriptstyle{A}$};
	\end{tikzpicture}
	\caption{Quiver theory on the torus constructed from the domain walls  
\eqref{eq:SpDW}, \eqref{eq:mixDW}, and \eqref{eq:rest_DW}. The 
diagonal lines are  glued according to the letters $a$, $b$, $c$, $d$ so that 
every $SO$ or $USp$ node connects to one $USp$, one $SO$, and two $SU$ nodes.}
	\label{fig:torusT}
\end{figure}
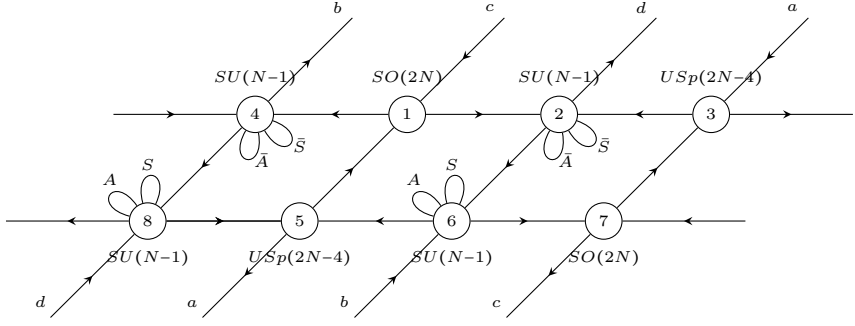

After consistently gluing all domain walls \eqref{eq:SpDW}, 
\eqref{eq:mixDW}, \eqref{eq:rest_DW},  and 
integrating all massive vertical fields, one has constructed the  
quiver theory displayed in Figure \ref{fig:torusT}, which would be understood 
as placed on a torus.
The choice of 
gluing in Figure \ref{fig:torusT} is not arbitrary, because the 
$\beta$-functions of the $USp$ and $SO$ nodes are zero up to leading order only 
on this way of connecting the lines. In more detail, one can straightforwardly 
compute, using  
\eqref{eq:NSVZ}, that
\begin{subequations}
\label{eq:beta-fcts}
\begin{align}
\beta_{SO}&=3\cdot 2(N-1)-\left(2(N-1)-2+2(N-1)+2+2(N-1) \right)+\mc 
O(g_{SO}^2) \notag\\*
&=0+\mc  O(g_{SO}^2) \,,\\
\beta_{USp}&=3\cdot 2(N-1)-\left(2(N-1)-2+2(N-1)+2+2(N-1) \right)+\mc 
O(g_{USp}^2) \notag \\*
&=0+\mc O(g_{USp}^2)\,.
\end{align}
\end{subequations}
The $\beta$-functions of the $SU$ gauge nodes are positive. 
From the perturbative analysis based on the $\beta$-functions, one would expect 
that the theory in Figure \ref{fig:torusT} becomes free in the IR.

Since all four domain walls are conformally glued, the 
$U(1)_R$ symmetry is preserved. Also, the $U(1)_R$-gravity anomaly 
is additive, so for the quiver of Figure \ref{fig:torusT} 
\begin{align}
\Tr\left(U(1)_R \right)=\Tr\left(U(1)_R^3 \right)=0
\end{align}
holds. Therefore, the central charge $a$ is zero in the UV regime. However, the 
RG flow will trigger the UV $U(1)_R$ symmetry to mix with other global $U(1)$ 
symmetries, such that one has to derive the genuine IR $R$-symmetry of 
the theory via $a$-maximisation \cite{Intriligator:2003jj}.
\paragraph{$\boldsymbol{a}$-maximisation.}
To begin with, one determines how many non-anomalous global $U(1)$ 
symmetries exists for the quiver. For bookkeeping, 
the 
gauge nodes are numbered, as shown in Figure \ref{fig:torusT}, such that 
matter fields are labeled by the nodes they 
connect, see Table \ref{tab:fields_charges} for details.

The matters fields carry charges under all compatible global symmetries that 
are `t~Hooft anomaly free for each gauge node as well as neutral with respect 
to the superpotentials. Recall that, besides regular 
quartic superpotential 
interactions, there are also superpotential terms for the (anti-)symmetric 
matter fields. For example, the fields $\bar{A}_2$ and $\bar{S}_2$ have 
superpotential as in \eqref{P2}, i.e.
\begin{align}
\mathcal{W}=q_{12}^2 \bar{S}_2 +q_{32}^2 \bar{A}_2\,,
\label{eq:superpot_anti-sym}
\end{align}
where all indices of these fields are omitted. Carefully evaluating all 
constraints shows that there are two $U(1)$ flavour symmetries with 
charges 
as displayed in Table \ref{tab:fields_charges}.
\begin{table}[t]
\centering
\begin{tabular}{c|cccccccc}
\toprule 
fields &$X_{71}$ & $X_{51}$ & $X_{53}$ & $X_{73}$ & $X_{26}$ & $X_{46}$ & 
$X_{48}$ & $X_{28}$ \\ \midrule
$U(1)_1$ &$0$ & $0$ & $0$ & $0$ &  $2$ & $0$ & $-2$ & $0$ \\
$U(1)_2$ & $-\alpha_-^2$ & $\alpha_+\alpha_-$ & $-\alpha_+^2$ & 
$\alpha_+\alpha_-$  & $-4-\alpha_+\alpha_-$ & $\alpha_+\alpha_-$ & 
$-\alpha_+\alpha_-$ & $\alpha_+\alpha_-$ \\ \midrule
%
fields & $q_{12}$ & $q_{32}$ & $q_{34}$ & $q_{14}$ & $q_{85}$ & $q_{65}$ & 
$q_{67}$ & $q_{87}$ \\ \midrule
$U(1)_1$ & $\alpha_-$ & $-\alpha_+$ & $\alpha_+$ & $-\alpha_-$ &  $\alpha_+$ & 
$-\alpha_+$ &  $\alpha_-$ & $-\alpha_-$ \\
$U(1)_2$ &  $-2\alpha_-$ & $2\alpha_+$ & $0$ & $0$ &   $0$ & $2\alpha_+$ &  
$-2\alpha_-$ & $0$ \\ \midrule
%
fields & $\bar{A}_2$ & $\bar{S}_2$ &  $\bar{A}_4$ & $\bar{S}_4$ & $A_6$ & $S_6$ 
&$A_8$ & $S_8$\\ 
\midrule
$U(1)_1$ &  $2\alpha_+$ & $-2\alpha_-$ &  $-2\alpha_+$ & $2\alpha_-$ & 
$2\alpha_+$ & $-2\alpha_-$ & $-2\alpha_+$ & 
$2\alpha_-$ \\
$U(1)_2$ & $-4\alpha_+$ & $4\alpha_-$ & $0$ & $0$  & $-4\alpha_+$ & $4\alpha_-$ 
& $0$ & $0$ \\
\bottomrule
\end{tabular}
\caption{Charge assignment of the fields in the quiver gauge theory 
of Figure \ref{fig:torusT}. Here, $\alpha_\pm=\frac{N-1}{2}\pm 1$.}
\label{tab:fields_charges}
\end{table}
Having derived the consistent charge assignment, one can proceed to compute 
the central charge in the IR regime, by defining a trial $R$-symmetry via
\begin{align}
R=U(1)_R+x_1 U(1)_1+x_2 U(1)_2\,.
\end{align}
Carrying out the $a$-maximisation leads to the solution
\begin{align}
x_1=x_2=\frac{2}{3(N-1)^2}\sqrt{\frac{10}{7}}\,, 
\label{eq:sol_a-max}
\end{align}
and the central charges $a$ and $c$ are then computed to be
\begin{align}
a=\frac{5}{3}\sqrt{\frac{10}{7}}
 \quad \text{and} \quad 
c=\frac{11}{6}\sqrt{\frac{10}{7}}  \,.
\end{align}
This central charge is positive, but $N$-independent. This is an unexpected 
behaviour from the $6$d point of view. 

Inspecting the solution \eqref{eq:sol_a-max}, it follows from Table 
\ref{tab:fields_charges} that for generic values of $N$
\begin{align}
\begin{aligned}
-\frac{5}{100} < &R_{\mathrm{IR}} \left(q_{14}, q_{12}, q_{67}, q_{87} \right) 
=  -\frac{1}{3}\sqrt{\frac{10}{7}}\frac{N-3}{(N-1)^2}
<0  \; , \\
 0< &R_{\mathrm{IR}} \left(q_{32}, q_{34}, q_{85}, q_{65} 
\right)=\frac{1}{3}\sqrt{\frac{10}{7}}\frac{N+1}{(N-1)^2} <\frac{2}{3} 
\;,
\end{aligned}
\label{eq:R-charge_horizontal}
\end{align}
while all other fields have IR $R$-charge above the unitarity bound. 
In particular, note that zero is a strict bound in 
\eqref{eq:R-charge_horizontal} for any finite $N$. 
The (anti-)symmetric chiral fields (and conjugates thereof) have IR 
$R$-charges that rapidly approach $2$ such that for $N\geq 10$, these are 
essentially equal to their UV $R$-charges.  
Nevertheless, the existence of fundamental fields with $R$-charge below the 
unitarity bound requires to carefully consider gauge 
invariant operators composed of the fields.
\paragraph{Gauge invariant operators and unitarity bound.} 
For the ring of gauge invariant chiral operators (GIO), one can, in principle, 
compute 
a generating function \cite{Hanany:2010zz} by an integral of the 
form\footnote{Note that it is not necessary to consider the chiral ring, i.e.\ 
one does not need to quotient by the F-term relations, if one just wants to find 
the generating set of gauge invariant chiral operators.}
\begin{align}
 F_{\mathrm{GIO}} = \prod_a \int \mu_{G_a} 
\mathrm{PE}\left[\sum_{\text{chirals}} \chi_{\mathrm{chirals}}^{ G} 
t^{R(\mathrm{chirals})}
\right]
\end{align}
where the integral is taken over the Haar measures of each group $G_a$ 
appearing in the gauge group $G=\prod_a G_a$ and the chirals contribute via 
their group characters $ \chi_{\mathrm{chirals}}^{ G}$.  Hence, $ 
F_{\mathrm{GIO}} $ becomes a Laurent series in $t$, the suitable UV R-charge 
fugacity. Assuming that the UV R-charges are non-negative, the ring of gauge 
invariant chiral operators is finitely generated. Consequently, it would be 
sufficient to analyse IR R-charges of the generating set of gauge invariant 
operators. However, due to the extensive nature of the quiver theory in Figure 
\ref{fig:torusT}, this approach is computationally challenging. Instead, an 
inspection of the quiver reveals the following basic gauge invariants:
\begin{itemize}
 \item GIOs that do not violate unitarity 
 \begin{compactenum}[(i)]
 \item There are gauge invariant operators which do not violate the unitarity 
bound. To name a few examples, consider closed loops in the quiver of Figure 
\ref{fig:torusT} that obey concatenation of arrows. 
The so-to-say smallest loops are all given by superpotential terms. By 
construction, the superpotential terms are of UV $R$-charge $2$ and singlets 
under the $U(1)_1$, $U(1)_2$ symmetries. Thus, these gauge-invariant operators 
have $R$-charge $2$ in the IR.
\item Gauge invariant operators corresponding to larger closed loops formed by 
concatenating arrows in the quiver, i.e.\ not given by a superpotential term, 
do not violate the unitarity bound either.  To see this, it is enough to notice 
how small the $R$-charges \eqref{eq:R-charge_horizontal} of the unitarity bound 
violating fields are, and to see that loops necessarily contain vertical chiral 
fields. The $R$-charges of these fields are large enough to compensate the 
small negative R-charges of $q_{14}$, $q_{12}$, $q_{67}$, or $q_{87}$ such that 
the resulting gauge invariant operator has $R$-charge above the unitarity 
bound. 
\item In addition, there are also gauge invariant operators that disobey 
concatenation of arrows; for instance, the $SO(2N)$ and $USp(2N{-}4)$ gauge 
nodes each admit invariant 2-tensors. These give rise to Meson-type operators 
\begin{align}
 \delta^{\alpha_7 \beta_7} \delta^{\alpha_1 \beta_1} (X_{71})_{\alpha_7 
\alpha_1} (X_{71})_{\beta_7 \beta_1}  
 \;, \quad
 \delta^{a_5 b_5} \delta^{a_3 b_3} (X_{53})_{a_5 a_3} (X_{53})_{b_5 b_3} \,,
\end{align}
where $\alpha,\beta$ are $SO(2N)$ indices on nodes $1$ and $7$, while 
$a,b$ are $USp(2N{-}4)$ indices on nodes $3$ and $5$. A calculation then 
proves that the $R$-charges of the Meson-type operators are above the unitarity 
bound.
On the other hand, one could consider the Pfaffian for an $SO(2N)$ gauge node 
which could be constructed as
\begin{align}
 \mathrm{Pf}\left[ (q_{67})_\alpha^{i} (A_6)_{[ij]} (q_{67})_\beta^{j} \right]
\end{align}
where $\alpha,\beta$ are $SO(2N)$ indices and $i,j$ are $SU(N{-}1)$ indices. 
The Pfaffian 
is then constructed for the antisymmetric $(\alpha \beta)$ indices. However, 
already the combination of the three chirals has $R$-charge above the 
unitarity bound.
\end{compactenum}
%
%
\item GIOs that do violate unitarity
\begin{compactenum}[(i)]
 \item There are Baryon-type operators constructed solely from field with 
negative IR $R$-charge, i.e.
\begin{subequations}
\label{eq:baryon_torus}
\begin{align}
 \mathcal{B}_{q_{14}q_{12}} = \varepsilon^{i_1 \ldots i_{N-1}} \varepsilon^{j_1 
\ldots j_{N-1}} &(q_{14})_{i_1 \alpha_1} \cdots (q_{14})_{i_{N-1} \alpha_{N-1}} 
\\
&\cdot 
(q_{12})_{j_1 \beta_1} \cdots (q_{12})_{j_{N-1} \beta_{N-1}}
\delta^{\alpha_1 \beta_1} \cdots \delta^{\alpha_{N-1} \beta_{N-1}} 
 \notag 
\\
\mathcal{B}_{q_{14}q_{14}} = \varepsilon^{i_1 \ldots i_{N-1}} \varepsilon^{j_1 
\ldots j_{N-1}} &(q_{14})_{i_1 \alpha_1} \cdots (q_{14})_{i_{N-1} \alpha_{N-1}} 
\\
&\cdot (q_{14})_{j_1 \beta_1} \cdots (q_{14})_{j_{N-1} \beta_{N-1}}
\delta^{\alpha_1 \beta_1} \cdots \delta^{\alpha_{N-1} \beta_{N-1}} 
\notag \\
\mathcal{B}_{q_{12}q_{12}} = \varepsilon^{i_1 \ldots i_{N-1}} \varepsilon^{j_1 
\ldots j_{N-1}} &(q_{12})_{i_1 \alpha_1} \cdots (q_{12})_{i_{N-1} \alpha_{N-1}} 
\\
&\cdot (q_{12})_{j_1 \beta_1} \cdots (q_{12})_{j_{N-1} \beta_{N-1}}
\delta^{\alpha_1 \beta_1} \cdots \delta^{\alpha_{N-1} \beta_{N-1}} 
\notag 
\end{align}
\end{subequations}
with the $SO(2N)$ indices $\alpha,\beta \in \{1,2,\ldots, 2N\}$ and the 
$SU(N{-}1)$ indices $i,j\in \{1,2,\ldots,N-1\}$. By the same 
reasoning, one constructs the GIOs $\mathcal{B}_{q_{67}q_{87}}$,  
$\mathcal{B}_{q_{67}q_{67}}$,  $\mathcal{B}_{q_{87}q_{87}}$. Besides these 
generalised Baryons, one may reconsider Pfaffian-type invariants of the form
\begin{align}
\label{eq:Pfaffian_torus}
\begin{aligned}
 \mathcal{P}_{q_{12}q_{14}} = 
 \mathrm{Pf}\bigg[ \big(
  \varepsilon^{i_1 \ldots i_{N-1}} \varepsilon^{j_1 
\ldots j_{N-1}} &(q_{14})_{i_1 \alpha_1} \cdots (q_{14})_{i_{N-1} 
\alpha_{N-1}}\\
&\cdot (q_{12})_{j_1 \beta_1} \cdots (q_{12})_{j_{N-1} \beta_{N-1}}
\delta^{\alpha_2 \beta_2} \cdots \delta^{\alpha_{N-1} \beta_{N-1}} 
\big)_{\alpha_1 \beta_1}
\bigg]
\end{aligned}
\end{align}
where the $SO(2N)$ Pfaffian is constructed from the anti-symmetric part in 
the $\alpha_1,\beta_1$ indices. Likewise, the Pfaffian 
$\mathcal{P}_{q_{67}q_{87}} $ can be constructed.
Then 
\eqref{eq:R-charge_horizontal} implies 
\begin{align}
R_{\mathrm{IR}}( 
\mathcal{B}_{q_{14}q_{12}},  
\mathcal{B}_{q_{14}q_{14}},  
\mathcal{B}_{q_{12}q_{12}}, 
\mathcal{B}_{q_{67}q_{87}},
\mathcal{B}_{q_{67}q_{67}},
\mathcal{B}_{q_{87}q_{87}},
\mathcal{P}_{q_{12}q_{14}},
\mathcal{P}_{q_{67}q_{87}}
) <0 \,.
\end{align}
\item Apart from gauge invariant operators constructed from fundamental chirals 
with negative $R$-charge, one can also consider operators with positive 
$R$-charge below the unitarity bound as long as enough $q_{12}$, $q_{14}$, 
$q_{67}$, $q_{87}$ are involved. As example, considered the following GIO:
\begin{subequations}
\begin{align}
\mathcal{O}_{q_{14}q_{12}}^{(1)}= J^{ab} 
(q_{34})_{i_1 a}  (q_{32})_{j_1 b} \varepsilon^{i_1 \ldots i_{N-1}} 
\varepsilon^{j_1 \ldots j_{N-1}}
 &(q_{14})_{i_2 \alpha_2} \cdots (q_{14})_{i_{N-1} \alpha_{N-1}} 
\\
&\cdot
(q_{12})_{j_2 \beta_2} \cdots (q_{12})_{j_{N-1} \beta_{N-1}}
\delta^{\alpha_2 \beta_2} \cdots \delta^{\alpha_{N-1} \beta_{N-1}} \notag 
\end{align}
which can be generalised to $l=1,2,\ldots,N-1$
\begin{align}
\mathcal{O}_{q_{14}q_{12}}^{(l)} =J^{a_1 b_1} 
(q_{34})_{i_1 a_1}  (q_{32})_{j_1 b_1} \cdots  J^{a_l b_l} 
&(q_{34})_{i_l a_l}  (q_{32})_{j_l b_l} \notag\\
\varepsilon^{i_1 \ldots i_{N-1}} 
\varepsilon^{j_1 \ldots j_{N-1}}
 &(q_{14})_{i_{l+1} \alpha_{l+1}} \cdots (q_{14})_{i_{N-1} \alpha_{N-1}} 
\\
&\cdot
(q_{12})_{j_{l+1} \beta_{l+1}} \cdots (q_{12})_{j_{N-1} \beta_{N-1}}
\delta^{\alpha_{l+1} \beta_{l+1}} \cdots \delta^{\alpha_{N-1} \beta_{N-1}} 
\notag 
\end{align}
\end{subequations}
with the $USp(2N{-}4)$ indices $a,b \in\{1,2,\ldots,2N-4\}$. 
Similarly, one also finds
\begin{subequations}%
\begin{align}%
\widetilde{\mathcal{O}}_{q_{14}q_{12}}^{(1)}= 
&J^{ab} (q_{34})_{i_1 a}  (q_{34})_{i_2 b}
\varepsilon^{i_1 \ldots i_{N-1}} 
(q_{14})_{i_3 \alpha_3} \cdots (q_{14})_{i_{N-1} \alpha_{N-1}} 
 \\
 &\cdot J^{cd}  (q_{32})_{j_1 c}  (q_{32})_{j_2 d}
 \varepsilon^{j_1 \ldots j_{N-1}}
 (q_{12})_{j_3 \beta_3} \cdots (q_{12})_{j_{N-1} \beta_{N-1}}
\delta^{\alpha_3 \beta_3} \cdots \delta^{\alpha_{N-1} \beta_{N-1}} \notag 
\end{align}
and generalisations thereof
\begin{align}
\widetilde{\mathcal{O}}_{q_{14}q_{12}}^{(l)}= 
&J^{a_1b_1} (q_{34})_{i_1 a_1}  (q_{34})_{i_2 b_1}
\cdots
J^{a_l b_l} (q_{34})_{i_{2l-1} a_l}  (q_{34})_{i_{2l} b_l} \\
&\cdot \varepsilon^{i_1 \ldots i_{N-1}} 
(q_{14})_{i_{2l+1} \alpha_{2l+1}} \cdots (q_{14})_{i_{N-1} \alpha_{N-1}} 
\notag \\
 &\cdot J^{c_1 d_1}  (q_{32})_{j_1 c_1}  (q_{32})_{j_2 d_1}
 \cdots 
 J^{c_l d_l}  (q_{32})_{j_{2l-1} c_l}  (q_{32})_{j_{2l} d_l} \notag\\
 &\cdot \varepsilon^{j_1 \ldots j_{N-1}}
 (q_{12})_{j_3 \beta_2} \cdots (q_{12})_{j_{N-1} \beta_{N-1}}
\delta^{\alpha_{2l+1} \beta_{2l+1}} \cdots \delta^{\alpha_{N-1} \beta_{N-1}} 
\,.
\notag 
\end{align}
\end{subequations}
From Table \ref{tab:fields_charges} and the solution \eqref{eq:sol_a-max}, one 
finds
\begin{subequations}
\begin{align}
 R_{\mathrm{IR}} \left(\mathcal{O}_{q_{14}q_{12}}^{(l)} \right)
 &= \frac{2}{3}\sqrt{\frac{10}{7}} \cdot \frac{3+2l-N}{N-1} <0 \quad \text{for 
} 
2l<N-3 \,,\\
 R_{\mathrm{IR}} \left(\widetilde{\mathcal{O}}_{q_{14}q_{12}}^{(l)} \right)
 &= \frac{2}{3}\sqrt{\frac{10}{7}} \cdot \frac{3+4l-N}{N-1} <0 \quad \text{for 
} 
4l<N-3 \,,
\end{align}
\end{subequations}
such that the $R$-charge is smaller than $\tfrac{2}{3}$ for large enough $N$.
Analogously, one defines the operators $\mathcal{O}_{q_{67}q_{87}}^{(l)}$ and 
$\widetilde{\mathcal{O}}_{q_{67}q_{87}}^{(l)}$.
\item As a last example, one can even construct unitarity violating operators 
by combining the $q_{12}$, $q_{14}$, $q_{67}$, $q_{87}$ with some of the 
vertical fields. For instance 
\begin{align}
\mathcal{O}=
&\varepsilon^{i_1 \ldots i_{N-1}} \varepsilon^{j_1 \ldots j_{N-1}}
(q_{14})_{i_{1} \alpha_{1}} \cdots (q_{14})_{i_{N-1} \alpha_{N-1}}
 (q_{12})_{j_1 \beta_1} \cdots (q_{12})_{j_{N-1} \beta_{N-1}}
\delta^{\alpha_{2} \beta_{2}} \cdots \delta^{\alpha_{N-1} \beta_{N-1}} \notag 
\\ 
&\cdot \varepsilon_{k_1 \ldots k_{N-1}} \varepsilon_{l_1 \ldots l_{N-1}}
(q_{67})^{k_{1}}_{\rho_{1}} \cdots (q_{67})^{k_{N-1}}_{\rho_{N-1}}
 (q_{87})^{l_1}_{\sigma_1} \cdots (q_{87})^{l_{N-1}}_{\sigma_{N-1}}
\delta^{\rho_{2} \sigma_{2}} \cdots \delta^{\rho_{N-1} \sigma_{N-1}}
\notag\\
&\cdot (X_{71})^{\alpha_1 \rho_1}(X_{71})^{\beta_1 \sigma_1}
\end{align}
which has $R$-charge given by
\begin{align}
  R_{\mathrm{IR}}\left(  \mathcal{O}\right) = 
  \frac{42-5\sqrt{70}}{21} +\frac{12\sqrt{70}}{21 (N-1)} - 
\frac{4\sqrt{70}}{21(N-1)^2}\,.
\end{align}
For large enough $N$, the operator $\mathcal{O}$ violates the unitarity bound.
\end{compactenum}
\end{itemize}
Consequently, GIOs below the unitarity bound need to be taken care of. 
Following the discussion in \cite{Kutasov:2003iy}, one would conclude that 
such gauge-invariant operators become free somewhere in the RG-flow and 
decouple. Hence, their contribution to the $a$-maximisation has to be subtracted 
in order to focus on the (potentially) interacting SCFT. As proposed in 
\cite{Benvenuti:2017lle}, this can be realised by introducing a 
\emph{flip field} $\beta_{\mathcal{O}}$  for a unitarity violating gauge 
invariant operator $\mathcal{O}$ such that
\begin{compactenum}[(i)]
 \item $\beta_{\mathcal{O}}$ is a gauge singlet, and
 \item $\beta_{\mathcal{O}}$ couples via the superpotential term $ 
\mathcal{W} 
\sim
\beta_{\mathcal{O}} \mathcal{O}$.
\end{compactenum}
Performing an a-maximisation with flip-fields for all the unitarity violating 
GIOs discussed above, leads to an IR $R$-charge assignment where also 
$X_{71}$, $X_{53}$, $X_{26}$, $X_{48}$ have negative $R$-charges. As a 
consequence, many more GIOs fall below the unitarity bound and would need to 
be taken care of. 

The appearing problem of unitarity-violating GIOs seems to be linked with the 
positivity of the $\beta$-functions \eqref{eq:beta-fcts}. The expectation is 
that the theory becomes free in the IR; therefore, it is suggestive to 
interpret the appearance of the large number of non-unitary GIOs at each 
iteration of the $a$-maximisation as indication that the process only 
terminates after all GIOs are removed from the theory. Hence, the theory would 
become free in the IR.
\paragraph{Multiple layers.}
The construction that led to the quiver theory of Figure \ref{fig:torusT} can 
be repeated to have more layers. In other words, gluing several copies along 
the legs labeled $a$, $b$, $c$, $d$. Thorough 
analysis of the anomaly and superpotential constraints for  
$k=2,3,4$ layers reveals that there exist three non-anomalous global $U(1)$ 
symmetries.
As a non-trivial consistency check of the calculation, one verifies that 
$\Tr(U(1)_R)=\Tr(U(1)_R^3)=0$ in the UV. 
Performing $a$-maximisation with respect to these three global $U(1)$ 
symmetries, one finds 
\begin{align}
 a_{k\text{ layers}} = k\cdot \frac{5}{3}\sqrt{\frac{10}{7}} 
 \quad \text{and} \quad 
 c_{k\text{ layers}} = k \cdot \frac{11}{6}\sqrt{\frac{10}{7}} \,.
\end{align}
As in the single layer case, one finds several gauge-variant chiral fields that 
violate unitarity. In more detail, essentially all horizontal chirals are 
below the unitarity; in particular, horizontal chirals transforming as 
bifundamentals of some $SU(N-1)$ and $SO(2N)$ have negative IR $R$-charge. 
Repeating the discussion of the 
single layer case reveals an increasingly large number of unitarity-violating 
GIOs 
which would need to be taken care of. Again, this is taken as indication that 
the multiple layer theory becomes free in the IR, in agreement with the 
expectation from the $\beta$-function analysis \eqref{eq:beta-fcts}. As a 
consistence check, the multiple layer analysis confirms the behaviour studied 
in detail in the single layer case.
%
%
%
\subsection{Quiver theory on 2-punctured sphere}
\label{sec:quiver_2-sphere+puncture}
Besides the theory on the torus, one can also attempt to study a meaningful 
$4$d theory on a two-punctured sphere. Analogously to the theory on the 
2-torus, two perspectives are taken: (i) the compactification of the $6$d 
theory \eqref{eq:6d_quiver_D-type} on a $2$-punctured sphere with fluxes, and 
(ii) construction via fundamental domain walls on the tube glued together with 
suitable punctures. 
\subsubsection{6d theory on 2-punctured sphere}
\label{sec:2-sphere_2-puncture}
Recall the anomaly 6-form \eqref{eq:4d_anomaly_punct-sphere} for the 
compactification on a 2-sphere with $s$ punctures and specialise to $s=2$. 
Proceeding to $a$-maximisation as in Section \ref{sec:a-max_6d}, one observes 
that there are no 
changes to the trace contributions which one reads off from the 
anomaly polynomial.
Hence, \eqref{eq:trace_terms_torus} are valid for the two-punctured 
2-sphere  and the $a$-maximisation proceeds as in the $T^2$ case. In 
particular, 
the solutions found in \eqref{eq:ex_T2_k=2}, 
\eqref{eq:ex_T2_k=3}, \eqref{eq:ex_T2_k=4} remain valid, but the $a$-central 
charge needs to be adjusted.
As elaborated in \cite{Razamat:2016dpl,Razamat:2018gro}, the puncture 
contribution is intimately linked to the $5$d theory resulting from putting the 
$6$d theory on a circle \cite{Hayashi:2015vhy}. 
The  $5$d \none theory \eqref{eq:5d_quiver_affine_A} is composed of $k$ 
$SO(2N{+}2)$ and $k$ $USp(2N{-}2)$ nodes and the additional 
anomaly inflow contributions modify $\Tr(R_{IR})$ and $\Tr(R_{IR}^3)$, i.e.
\begin{align}
 \Tr(R_{\mathrm{pnct.}} ) = \Tr(R^3_{\mathrm{pnct.}} ) &= - \frac{s}{2} \cdot k 
\cdot  \left( \dim( SO(2N{+}2) )  + \dim( USp(2N{-}2) ) \right) \notag \\
&= -s\cdot k(2N^2-1) \,, \\
\Rightarrow \quad 
a_{S^2_s} &= a_{T^2} - \frac{3s}{16} k(2N^2+1) \,,
\end{align}
which relies on the assumption that the maximal symmetry in $5$d is realised.

On the other hand, if the punctures do not exhibit the full symmetry, but 
subgroups $H_{SO} \subset SO(2N{+}2)$ and $H_{Sp} \subset USp(2N{-}2)$, 
respectively, then the boundary conditions are chosen as in 
\eqref{eq:non-max_bc_puncture}. The anomaly inflow contributions, which 
originate from the $5$d $U(1)_R$ CS-term for the non-trivial $H_{Sp,SO}$ 
components of the adjoint chiral, change accordingly to
\begin{align}
 \Tr(R_{\mathrm{pnct.} } ) 
 = \Tr(R^3_{\mathrm{pnct.} } ) 
 = - \frac{s}{2} \cdot k 
\cdot  \left( \dim( H_{SO} )  + \dim( H_{Sp} ) \right) \,.
\end{align}
Suppose $H_{SO,Sp}$ are the maximal subgroups preserving the rank, $H_{SO}= 
SU(N{+}1)$ and $H_{Sp}=SU(N{-}1)$, then one finds
\begin{align}
 \Tr(R_{\mathrm{pnct.}} ) &= \Tr(R^3_{\mathrm{pnct.}} ) = 
 -s\cdot kN^2 \\
\Rightarrow \quad 
 a_{S^2_s} &= a_{T^2} - \frac{3s}{16} kN^2 \,.
\end{align}
%
%
\subsubsection{Quiver theory from domain walls}
\begin{figure}[t]
 \centering
 	\begin{tikzpicture}
 	\tikzset{node distance = 2cm}
	\tikzstyle{gauge} = [circle, draw,inner sep=1.5pt];
	\tikzstyle{flavour} = [regular polygon,regular polygon sides=4,inner 
sep=1pt, draw];
%
\node (g1) [gauge,label=below right:{$\scriptscriptstyle{SU(N{-}1)}$}] 
{$\scriptscriptstyle{1}$};
\node (g2) [gauge,right of=g1, label=below 
right:{$\scriptscriptstyle{SO(2N)}$}]{$\scriptscriptstyle{2}$};
\node (g3) [gauge,right of=g2,label=below 
right:{$\scriptscriptstyle{SU(N{-}1)}$}]{$\scriptscriptstyle{3}$};
\node (g4) [gauge,below of=g1,label=below 
right:{$\scriptscriptstyle{USp(2N{-}4)}$}] {$\scriptscriptstyle{4}$};
\node (g5) [gauge,right of=g4,label=below 
right:{$\scriptscriptstyle{SU(N{-}1)}$}] 
{$\scriptscriptstyle{5}$};
\node (g6) [gauge,right of=g5,label=below right:{$\scriptscriptstyle{SO(2N)}$}] 
{$\scriptscriptstyle{6}$};
\node (g7) [gauge,below of=g4,label=below 
right:{$\scriptscriptstyle{SU(N{-}1)}$}]{$\scriptscriptstyle{7}$};
\node (g8) [gauge,right of=g7,label=below 
right:{$\scriptscriptstyle{SO(2N)}$}]{$\scriptscriptstyle{8}$}
;
\node (g9) [gauge,right of=g8,label=below 
right:{$\scriptscriptstyle{SU(N{-}1)}$}] 
{$\scriptscriptstyle{9}$}; 
\node (g10) [gauge,below of=g7,label=below 
right:{$\scriptscriptstyle{USp(2N{-}4)}$}] {$\scriptscriptstyle{10}$};
\node (g11) [gauge,right of=g10,label=below 
right:{$\scriptscriptstyle{SU(N{-}1)}$}] {$\scriptscriptstyle{11}$};
\node (g12) [gauge,right of=g11,label=below 
right:{$\scriptscriptstyle{SO(2N)}$}]{$\scriptscriptstyle{12}$}
;
\node (L1) [flavour,left of=g1, 
label=left:{$\scriptscriptstyle{USp(2N{-}4)}$}] {$\scriptscriptstyle{L_1}$};
\node (L2) [flavour,left of=g4, 
label=left:{$\scriptscriptstyle{SU(N{-}1)}$}]{$\scriptscriptstyle{L_2}$};
\node (L3) [flavour,left 
of=g7,label=left:{$\scriptscriptstyle{USp(2N{-}4)}$}]{$\scriptscriptstyle{L_3}$}
;
\node (L4) [flavour,left 
of=g10,label=left:{$\scriptscriptstyle{SU(N{-}1)}$}] 
{$\scriptscriptstyle{L_4}$};
\node (R1) [flavour,right of=g3, 
label=right:{$\scriptscriptstyle{USp(2N{-}4)}$}] {$\scriptscriptstyle{R_1}$};
\node (R2) [flavour,right of=g6, 
label=right:{$\scriptscriptstyle{SU(N{-}1)}$}]{$\scriptscriptstyle{R_2}$};
\node (R3) [flavour, right 
of=g9,label=right:{$\scriptscriptstyle{USp(2N{-}4)}$}]{$\scriptscriptstyle{R_3}$
};
\node (R4) [flavour, right 
of=g12,label=right:{$\scriptscriptstyle{SU(N{-}1)}$}] 
{$\scriptscriptstyle{R_4}$};
%
\tikzset{node distance = 1cm}
\node(up1) [above of= L1] {};
\node(up2) [right of= up1 ,label=below:{$\scriptscriptstyle{a}$}] {};
\tikzset{node distance = 2cm}
\node(up3) [right of= up2,label=below:{$\scriptscriptstyle{b}$}] {};
\node(up4) [right of= up3,label=below:{$\scriptscriptstyle{c}$}] {};
\node(up5) [right of= up4,label=below:{$\scriptscriptstyle{d}$}] {};
\tikzset{node distance = 1cm}
\node(up6) [right of= up5] {};
\node(down1) [below of= L4] {};
\node(down2) [right of= down1 ,label=above:{$\scriptscriptstyle{a}$}] {};
\tikzset{node distance = 2cm}
\node(down3) [right of= down2,label=above:{$\scriptscriptstyle{b}$}] {};
\node(down4) [right of= down3,label=above:{$\scriptscriptstyle{c}$}] {};
\node(down5) [right of= down4,label=above:{$\scriptscriptstyle{d}$}] {};
\tikzset{node distance = 1cm}
\node(down6) [right of= down5] {};
    \begin{scope}[decoration={markings,mark =at position 0.5 with 
{\arrow{stealth}}}]
    \draw[postaction={decorate}] (L1) -- (up1);
    \draw[postaction={decorate}] (L2) -- (L1);
    \draw[postaction={decorate}] (L3) -- (L2);
    \draw[postaction={decorate}] (L4) -- (L3);
    \draw[postaction={decorate}] (down1) -- (L4);
    \draw[postaction={decorate}] (up6) -- (R1);
    \draw[postaction={decorate}] (R1) -- (R2);
    \draw[postaction={decorate}] (R2) -- (R3);
    \draw[postaction={decorate}] (R3) -- (R4);
    \draw[postaction={decorate}] (R4) -- (down6);
    \draw[postaction={decorate}] (L1) -- (g1);
    \draw[postaction={decorate}] (g2) -- (g1);
    \draw[postaction={decorate}] (g2) -- (g3);
    \draw[postaction={decorate}] (R1) -- (g3);
    \draw[postaction={decorate}] (L2) -- (g4);
    \draw[postaction={decorate}] (g5) -- (g4);
    \draw[postaction={decorate}] (g5) -- (g6);
    \draw[postaction={decorate}] (R2) -- (g6);
    \draw[postaction={decorate}] (L3)--(g7);
    \draw[postaction={decorate}] (g8)--(g7);
    \draw[postaction={decorate}] (g8)--(g9);
    \draw[postaction={decorate}] (R3) -- (g9);
    \draw[postaction={decorate}] (L4)--(g10);
    \draw[postaction={decorate}] (g11)--(g10);
    \draw[postaction={decorate}] (g11)--(g12);
    \draw[postaction={decorate}] (R4)--(g12);
    \draw[postaction={decorate}] (up2)--(L1);
    \draw[postaction={decorate}] (g1)--(up3);
    \draw[postaction={decorate}] (up4)--(g2);
    \draw[postaction={decorate}] (g3)--(up5);
    \draw[postaction={decorate}] (g1)--(L2);
    \draw[postaction={decorate}] (g4)--(g2);
    \draw[postaction={decorate}] (g3)--(g5);
    \draw[postaction={decorate}] (g6)--(R1);
    \draw[postaction={decorate}] (g4)--(L3);
    \draw[postaction={decorate}] (g7)--(g5);
    \draw[postaction={decorate}] (g6)--(g8);
    \draw[postaction={decorate}] (g9)--(R2);
    \draw[postaction={decorate}] (g7)--(L4);
    \draw[postaction={decorate}] (g10)--(g8);
    \draw[postaction={decorate}] (g9)--(g11);
    \draw[postaction={decorate}] (g12)--(R3);
    \draw[postaction={decorate}] (g10)--(down2);
    \draw[postaction={decorate}] (down3)--(g11);
    \draw[postaction={decorate}] (g12)--(down4);
    \draw[postaction={decorate}] (down5)--(R4);
     \end{scope}
%
\draw (g1)  to[in=-190,out=-230,loop] (g1);
\draw (g1)  to[in=-250,out=-290,loop] (g1);
\draw (-0.1,0.6) node {$\scriptscriptstyle{\bar{S}}$};
\draw (-0.6,0.3) node {$\scriptscriptstyle{\bar{A}}$};
\draw (g3)  to[in=-190,out=-230,loop] (g3);
\draw (g3)  to[in=-250,out=-290,loop] (g3);
\draw (-0.1+4,0.6) node {$\scriptscriptstyle{\bar{S}}$};
\draw (-0.6+4,0.3) node {$\scriptscriptstyle{\bar{A}}$};
\draw (g7)  to[in=-190,out=-230,loop] (g7);
\draw (g7)  to[in=-250,out=-290,loop] (g7);
\draw (-0.1,0.6-4) node {$\scriptscriptstyle{\bar{S}}$};
\draw (-0.6,0.3-4) node {$\scriptscriptstyle{\bar{A}}$};
\draw (g9)  to[in=-190,out=-230,loop] (g9);
\draw (g9)  to[in=-250,out=-290,loop] (g9);
\draw (-0.1+4,0.6-4) node {$\scriptscriptstyle{\bar{S}}$};
\draw (-0.6+4,0.3-4) node {$\scriptscriptstyle{\bar{A}}$};
\draw (g5)  to[in=-190,out=-230,loop] (g5);
\draw (g5)  to[in=-250,out=-290,loop] (g5);
\draw (-0.1+2,0.6-2) node {$\scriptscriptstyle{S}$};
\draw (-0.6+2,0.3-2) node {$\scriptscriptstyle{A}$};
\draw (g11)  to[in=-190,out=-230,loop] (g11);
\draw (g11)  to[in=-250,out=-290,loop] (g11);
\draw (-0.1+2,0.6-6) node {$\scriptscriptstyle{S}$};
\draw (-0.6+2,0.3-6) node {$\scriptscriptstyle{A}$};
\draw (L2)  to[in=-210,out=-250,loop] (L2);
\draw (-0.6-2,0.3-2) node {$\scriptscriptstyle{A}$};
\draw (L4)  to[in=-210,out=-250,loop] (L4);
\draw (-0.6-2,0.3-6) node {$\scriptscriptstyle{A}$};
\draw (R2)  to[in=-210,out=-250,loop] (R2);
\draw (-0.6+6,0.3-2) node {$\scriptscriptstyle{S}$};
\draw (R4)  to[in=-210,out=-250,loop] (R4);
\draw (-0.6+6,0.3-6) node {$\scriptscriptstyle{S}$};
	\end{tikzpicture}
\caption{A quiver theory on a two-punctures sphere, for $k=2$. The theory is 
constructed from the domain walls \eqref{eq:SpDW}, 
\eqref{eq:mixDW}, and \eqref{eq:rest_DW}, but it is not glued to a torus. As 
for the torus theory of Figure \ref{fig:torusT}, the diagonal lines are 
identified according to the labels $a$, $b$, $c$, $d$.}
	\label{fig:quiver_open}
\end{figure}
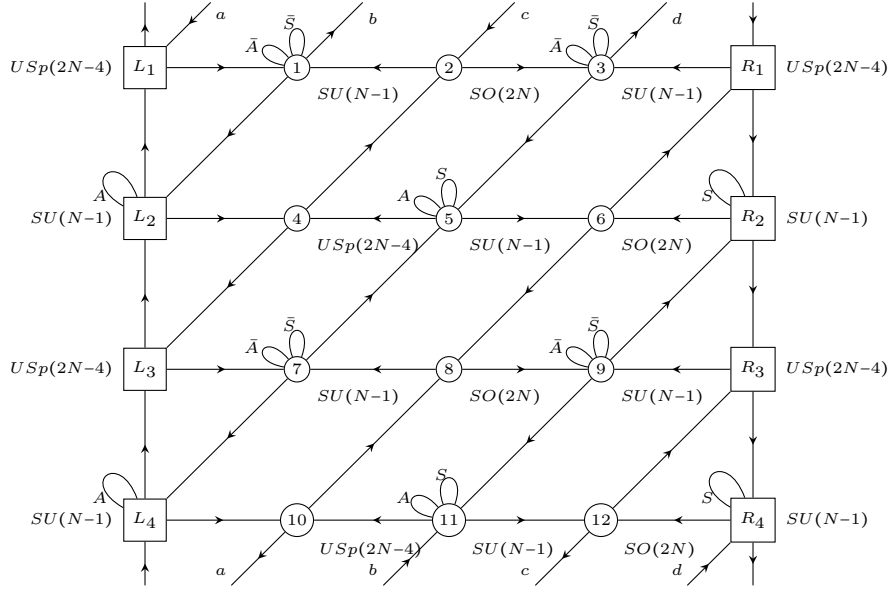

\begin{figure}[t]
\centering
\begin{center}
 \includegraphics[width=0.65\textwidth]{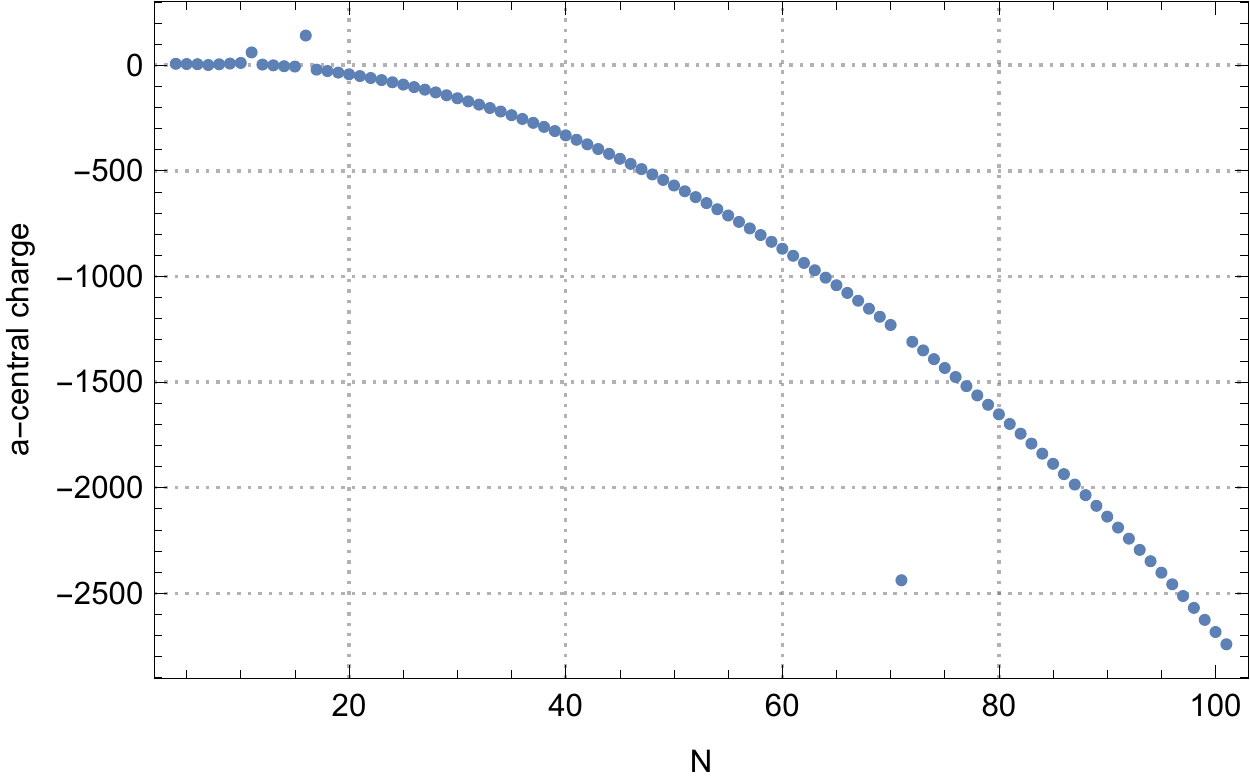}
 \end{center}
\caption{The numerical results of the $a$-maximisation:  the 
central charge is displayed against varying rank $N$ of the gauge nodes. Note 
that the numerical results for $N=12,16,71$ seem to be flawed.}
\label{fig:num_results}
\end{figure}

As an example, consider the theory in Figure 
\ref{fig:quiver_open} where the two punctures each have symmetry $USp(2N{-}4)^2 
\times SU(N{-}1)^2$.
From the quiver diagram, one straightforwardly 
derives that there are $8$ anomaly-free $U(1)$ flavour symmetries. Subsequently 
performing $a$-maximisation 
with respect to all of them leads to an $a$-central charge behaving as 
displayed 
in Figure \ref{fig:num_results}. 
\paragraph{Gauge invariant operators and unitarity bound.}
After the $a$-maximisation of the theory in Figure \ref{fig:quiver_open}, 
one has to inspect the IR $R$-charges again. Similar to the theory on the 
torus, one can summarise the numerical results as follows:
\begin{subequations}
\begin{align}
-\frac{1}{2} \leq &R_{IR} \left(q_{21}, q_{23}, q_{87}, q_{89} \right) <0 
\,,\quad 
 0 \leq
 R_{IR} \left( \text{other horizontal }q  \right) < \frac{2}{3} \;,\\
 &R_{IR} \left( \text{(anti-)symmetric chirals } A,S  \right) >1 \;,\\
 &R_{IR} \left( \text{boundary fields } Y  \right) >\frac{2}{3} \;,\\
 \frac{1}{2} \leq&R_{IR} \left(  X_{6R_1} , X_{12 R_3} , X_{4 L_3}, X_{10 L_1}  
\right) <\frac{2}{3} \;,\quad 
R_{IR} \left( \text{other vertical }  X \right) >\frac{2}{3} \,.
\end{align}
\end{subequations}
Again, the existence of gauge variant objects violating unitarity by itself is 
not problematic, but one needs to consider all gauge invariant operators 
carefully.
As in the Section \ref{sec:quiver_torus}, it is sufficient to focus on 
invariants build from 
unitarity violating chirals. It is immediate that some unitarity violating 
operators of the quiver theory in Figure \ref{fig:torusT} are also unitarity 
violating GIOs in the theory of Figure \ref{fig:quiver_open}; for instance, 
the analogs of the Baryon-type operators \eqref{eq:baryon_torus} and 
Pfaffian-type operators \eqref{eq:Pfaffian_torus} for $q_{21}$, $q_{23}$, 
$q_{87}$, $q_{89}$ do exist.
In addition, compared to the quiver on the torus of Figure 
\ref{fig:torusT}, the conceptually new gauge invariant operators are the 
following:
\begin{compactenum}[(i)]
 \item Since the quiver has flavour nodes, some gauge invariant operators can 
originate from paths starting and ending on a flavour node.
For instance,
\begin{align}
  (X_{7 L4})_{i}^{\ m_7} (q_{87})_{a m_7} \delta^{ab} (q_{89})_{b n_9} (X_{9 
R_2})_{j}^{\ n_9}
\end{align}
where the $SO(2N)$ gauge indices are contracted by the invariant tensor of 
$SO$. However, this operator does not violate unitarity. More generally, paths 
flavour nodes that originate from concatenation of arrows doe not introduce 
unitarity violating operators, because the $R$-charges of the 
involved horizontal and (anti-)symmetric chirals are large enough. 
\item Meson-type operators for horizontal chirals that transform under an 
$SO(2N)$ 
gauge node and an $SU(N{-}1)$ flavour node lead to operators violating the 
unitarity bound. For instance, 
\begin{subequations}
\label{eq:quiver_open_meson}
\begin{align}
\mathcal{M}_{R_2}^{(i,j)} =  \sum_{a,b=1}^{2N} \delta^{ab} (q_{R_2  6})_{a}^{i} 
(q_{R_2  6})_{b}^{j} \; , \quad 
 \mathcal{M}_{R_4}^{(i,j)} =  \sum_{a,b=1}^{2N} \delta^{ab} (q_{R_4  
12})_{a}^{i} 
(q_{R_4  12})_{b}^{j} \;, 
  \end{align}
which both transform in the 2nd-rank symmetric product of the anti-fundamental 
representation of an $SU(N-1)$ flavour group.
Similarly, Mesons constructed from chirals that transform 
under an $USp(2N{-}4)$ 
gauge node and an $SU(N{-}1)$ flavour node lead to problematic gauge invariant 
operators.
\begin{align}
\mathcal{M}_{L_2}^{[i,j]} =  \sum_{\alpha,\beta=1}^{2N-4} J^{\alpha \beta} 
(q_{L_2  
4})_{\alpha}^{i} 
(q_{L_2  4})_{\beta}^{j} \; , \quad 
 \mathcal{M}_{L_4}^{[i,j]} =  \sum_{\alpha,\beta=1}^{2N-4} J^{\alpha \beta} 
(q_{L_4  
10})_{\alpha}^{i} 
(q_{L_4  10})_{\beta}^{j} \;, 
  \end{align}
  \end{subequations}
  which transform in the 2nd-rank anti-symmetric product of the 
anti-fundamental representation of some $SU(N{-}1)$ flavour group.
\end{compactenum}
Again, the non-unitary GIOs are expected to become free somewhere on the 
RG-flow and decouple \cite{Kutasov:2003iy}. Then the $a$-maximisation is 
modified by suitable flip fields \cite{Benvenuti:2017lle}.
Inspecting the theory in Figure \ref{fig:quiver_open} reveals that 
(anti-)symmetric chirals 
$S_{R_6}$, $S_{R_2}$, $A_{L_2}$, and $A_{L_4}$ of the $SU(N{-}1)$ flavour nodes 
serves as flip fields for the Meson operators 
\eqref{eq:quiver_open_meson}, because of couplings like 
\eqref{eq:superpot_anti-sym}. Since, the flip fields have been included in the 
$a$-maximisation, the Meson-type operators are already removed. 
However, one still has to take care of all the non-unitary GIOs that descend 
from the quiver on the torus. Analogously to the discussion in Section 
\ref{sec:quiver_torus}, 
introducing all relevant flip fields, and re-perform $a$-maximisation is 
expected to yield even more GIOs below the unitarity bound, such that the 
theory in Figure \ref{fig:quiver_open}
is anticipated to become free in the IR.
%
%
%
\section{Conclusions}
\label{sec:conclusion}
In this paper we explored compactifications of $6$d $D$-type $(1,0)$ SCFTs on 
the torus and to some extent on the two-punctured sphere preserving $4$d 
$\mathcal{N}=1$ superconformal symmetry. We explicitly computed the $4$d 
central charges by integrating the anomaly polynomial of the $6$d theory turning 
on fluxes and determined their values in the deep IR using $a$-maximisation. We 
observe an interesting $N^{3\slash2}$ growth of these central charges where 
$N+1$ is 
the rank of our $D$-type quiver. The second part of the paper dealt with 
explicit constructions of candidate $4$d UV Lagrangians arising from such $6$d 
compactifications. Here we found that anomaly considerations at the level of the 
$4$d Lagrangian theory impose severe constraints on the gauge nodes and the 
matter content. We find that within our framework only non-maximal boundary 
conditions turn out to be consistent. 
In particular, we note that imposing UV $R$-charge assignments determined by 
$5$d domain wall constructions uniquely singles out non-maximal boundary 
conditions, which are novel and should be explored further.
However, the constructed $4$d quiver theories appear to be IR free, which is 
indicated from two directions. On the one hand, the $\beta$-functions are 
positive. On the other hand, the $a$-maximisation is spoiled by a large 
number of unitarity-violating gauge invariant operators. We expect that the 
process of introducing flip fields and re-performing $a$-maximisation only 
terminates once the free IR theory is reached.

In view of the $6$d prediction, the mismatch is not utterly surprising. 
Firstly, boundary conditions for orthogonal and symplectic gauge groups have not 
been thoroughly studied, yet. Secondly, non-maximal boundary conditions have not 
been explored from the point of view of anomalies even for $6$d parent 
theories of $A$-type. Thirdly, the anomaly 
polynomial has no information on potentially unitarity-violating operators or 
free operators. We expect that accounting correctly for these effects will 
correct the $N^{3\slash 2}$-dependence of the $a$-central charge in $4$d.
 
We postpone a more detailed study for future work.
\paragraph{Acknowledgements.}
We are grateful to Shlomo S.\ Razamat for discussions and useful suggestions in 
the early stage of this work. 
We would like to express our gratitude towards the referee whose thorough 
comments helped to improve this paper.
The research of J.C.\ is supported in part by the Chinese Academy of Sciences 
(CAS) Hundred-Talent Program and by Projects No.\ 11747601 and No.\ 11535011 
supported by National Natural Science Foundation of China.
The work of B.H.\ and M.S.\ was supported by the National 
Thousand-Young-Talents 
Program of China.
%
%
\begin{appendix}
\section{Notations, anomalies and \texorpdfstring{$\beta$}{beta}-functions}
\label{app:anomalies}
This appendix provides a summary of the group theoretical notations, 
and a brief review of various `t~Hooft anomalies and NSVZ 
$\beta$-functions. Following for instance \cite{Yamatsu:2015npn}, all required 
group theoretical constants and conventions are collected in 
Table \ref{tab:conventions_groups}.
For convenience, the formulae for various anomalies are recalled.
\paragraph{$\Tr\,(G^3)$ cubic anomaly.}
Since only gauge groups with matter fields in complex representations can have 
a cubic gauge anomaly, one only needs to consider $G=SU(N)$.
Thus, matter fields in the representation $\oplus_i n_i R_i$ contribute to 
cubic anomaly as follows
\begin{equation}
\Tr\,(SU(N)^3)=\sum_i n_i A(R_i)\,,
\label{cubic} 
\end{equation}
where $n_i$ is the multiplicity of the representation of $R_i$.
\paragraph{$\Tr\,(U(1)_R\,G^2)$ anomaly.} 
Contrary to the cubic gauge anomaly, all gauge groups potentially 
suffer from the mixed anomaly $\Tr\,(U(1)_R\,G^2)$. Since both, gauge and 
matter 
multiplets contribute, the anomaly can be evaluated via
\begin{equation}
\Tr\,(U(1)_R \,G^2)=t_2({\rm adj.})+\sum_i n_i\,(r_i-1)\,t_2(R_i)\,,
\label{RG2}
\end{equation}
where the first term is due to the gauginos ($R$-charge 1), and the second term 
accounts for chiral superfields of $R$-charge $r_i$, transforming in a
representation $R_i$.
\paragraph{$\Tr\,(U(1)_R)$ and $\Tr\,(U(1)_R^3)$ anomalies.} 
Lastly, the $\Tr\,(U(1)_R)$ and $\Tr\,(U(1)_R^3)$ anomalies can be evaluated via
\begin{align}
\begin{aligned}
\Tr\,\left(U(1)_R\right)&=\sum_\alpha  d_{G_\alpha}+\sum_a d_a (r_a-1) \,,\\
\Tr\,\left(U(1)_R^3\right)&=\sum_\alpha  d_{G_\alpha}+\sum_a d_a (r_a-1)^3 \,,
\end{aligned}
\label{eq:RR3}
\end{align}
where $\alpha$ and $a$ run over all gauge and chiral matter fields (with 
$R$-charge $r_a$), respectively. The multiplicity $d_a = \dim 
R_{\mathrm{gauge}} 
\cdot \dim R_{\mathrm{flavour}}$ of the $a$-th
chiral supermultiplet accounts for the dimensions with respect to the 
gauge and flavour symmetry representation.
\begin{table}[t]
\centering
\begin{tabular}{c|ccc}
\toprule
	 $G$ & $SU(N)$ & $USp(2N{-}2)$ & $SO(2N{+}2)$\\ \midrule
	 $r_G$ & $N-1$ & $N-1$ & $N+1$ \\
	 $h_G^\vee$ & $N$ & $N$ & $2N$ \\
	 $d_G$ & $N^2-1$ & $(N-1)(2N-1)$ & $(N+1)(2N+1)$ \\
	 $t_2({\rm fund.})$ & $1$ & 1 & $1$ \\
	 $t_2({\rm adj.})$ & $2N$ & $2N$ & $2N$ \\
	 $t_2({\rm sym.})$ & $N+2$ & -- & -- \\
	 $t_2({\rm antisym.})$ & $N-2$ & -- & -- \\

	 $A({\rm fund.})$ & $1$ & $0$ & $0$ \\
	 $A({\rm sym.})$ & $N+4$ & $0$ & $0$ \\
	 $A({\rm antisym.})$ & $N-4$ & $0$ & $0$ \\
	\bottomrule
\end{tabular}
\caption{Group theoretical constants. Here, $r_G$ denotes for the rank, $d_G$ 
is the dimension of $G$, and $h_G^\vee$ is the dual Coxeter number. For a 
given representation of a group $G$, $t_2({\rm rep.})$ denotes the second 
Dynkin index, while $A({\rm rep.})$ is the cubic anomaly coefficient.
The conventions for the second Dynkin index of $SU$ and $USp$ groups are 
adjusted to be the same as that of $SO$ groups by multiplying an additional 
factor of $2$.
}
\label{tab:conventions_groups}
\end{table}
\paragraph{NSVZ $\beta$-function.}
The \none NSVZ $\beta$-function of a gauge node is given by
\begin{equation}
\beta\left(\frac{8\pi^2}{g_c^2}\right)\equiv \mu\frac{\partial}{\partial 
\mu}\frac{8\pi^2}{g_c^2}=\frac{3t_2({\rm adj.})-\sum_i 
t_2(R_i)(1-\gamma_i(g_c))}{1-\frac{t_2({\rm adj.})}{8\pi^2}g_c^2}\,,
\end{equation}
where $\mu$ is the energy scale, and $\gamma_i(g_c)$'s are the anomalous 
dimensions of matter fields in representation $R_i$. Since in this work only 
the sign of the $\beta$-function is relevant, whose denominator is always 
positive for small enough $g_c^2$, one may solely focus on the numerator. 
Furthermore, the anomalous dimensions are proportional to $g_c^2$, which is a 
 second order corrections; thus, for an asymptotically free theory, one 
requires 
that the leading order contribution of 
$\beta\left(\frac{8\pi^2}{g_c^2}\right)$ must be non-negative, i.e.
\begin{align}
\beta\left(\frac{8\pi^2}{g_c^2}\right)\bigg\vert_{\rm leading\ order}\propto 
\left(3t_2({\rm adj.})-\sum_i t_2(R_i)\right)\geq 0\,,
\label{eq:NSVZ}
\end{align}
where $3t_2({\rm adj.})$ originates from the gauge multiplet, and the remainder 
accounts for the matter multiplets.
In fact, the condition \eqref{eq:NSVZ} is sufficient to verify if a 
theory is asymptotically or IR free, because \eqref{eq:NSVZ} yields the 
upper bound of the conformal window of the gauge node. For example, consider a  
$SU(N_c)$ gauge node with $N_f$ quarks $(Q\,,\tilde Q)_i$ in the 
(anti-)fundamental representation, then \eqref{eq:NSVZ} implies that
\begin{align}
3\cdot 2N_c-2N_f\cdot 1\geq 0 \quad \Leftrightarrow \quad  N_f\leq 3N_c\,,
\end{align}
which is exactly the upper bound of the conformal window for a $SU(N_c)$ gauge 
node. Likewise, the statement is valid for $SO(N)$ and 
$USp(2N)$ gauge nodes too. For instance, consider a $USp(2N_c)$ gauge 
node with $2N_f$ chiral fields $Q_i$ in the fundamental 
representation of $USp(2N_c)$ group. Then \eqref{eq:NSVZ} yields
\begin{align}
3\cdot (2N_c+2)-2N_f\cdot 1\geq 0 \quad \Leftrightarrow \quad N_f\leq 
3(N_c+1)\,,
\end{align}
and, again, $3(N_c+1)$ is precisely the upper bound of the conformal 
window for a $USp(2N_c)$ gauge group. For a $SO(2N_c)$ gauge group with $N_f$ 
chiral fields $Q_i$ in the vector representation of $SO(2N_c)$, one 
can verify that the upper bound is given by $3(N_c-1)$.
\end{appendix}

 \bibliographystyle{JHEP}     
 {\footnotesize{\bibliography{references}}}
\end{document}